\documentstyle[12pt,epsf,axodraw]{article}
\newcommand{\la}[1]{\label{#1}}
\newcommand{\be}{\begin{equation}}
\newcommand{\ee}{\end{equation}}
\newcommand{\ba}{\begin{eqnarray}}
\newcommand{\ea}{\end{eqnarray}}
\newcommand{\bi}{\begin{itemize}}
\newcommand{\ei}{\end{itemize}}
\newcommand{\rmi}[1]{{\mbox{\scriptsize #1}}}

\newcommand{\nr}[1]{(\ref{#1})}
\newcommand{\tr}{{\rm Tr\,}}
\newcommand{\nn}{\nonumber}
\newcommand{\fr}[2]{{\frac{#1}{#2}}}
\newcommand{\msbar}{\overline{\mbox{\rm MS}}}

\newcommand{\bfx}{{\bf x}}
\newcommand{\mH}{m_{0^{++}}}
\newcommand{\tE}{E} 
\newcommand{\tM}{M} 
\newcommand{\eq}{Eq.\,}
\newcommand{\eqs}{Eqs.\,}

\def\lsi{\raise0.3ex\hbox{$<$\kern-0.75em\raise-1.1ex\hbox{$\sim$}}}
\def\gsi{\raise0.3ex\hbox{$>$\kern-0.75em\raise-1.1ex\hbox{$\sim$}}}
\newcommand{\lsim}{\mathop{\lsi}}
\newcommand{\gsim}{\mathop{\gsi}}

\begin{document}

\begin{titlepage}
\begin{flushright}
CERN-TH/98-08\\
NORDITA-98/30HE\\
hep-lat/9805013\\
\end{flushright}
\begin{centering}
\vfill

{\bf THE UNIVERSALITY CLASS OF THE ELECTROWEAK THEORY}
\vspace{0.8cm}

K. Rummukainen$^{\rm a}$\footnote{kari@nordita.dk},
M. Tsypin$^{\rm b}$\footnote{tsypin@td.lpi.ac.ru},
K. Kajantie$^{\rm c,d}$\footnote{keijo.kajantie@cern.ch},
M. Laine$^{\rm c,d}$\footnote{mikko.laine@cern.ch}, \\
 and
M. Shaposhnikov$^{\rm c}$\footnote{mshaposh@nxth04.cern.ch} \\

\vspace{0.3cm}
{\em $^{\rm a}$Nordita, Blegdamsvej 17, DK-2100 Copenhagen,
Denmark\\}
\vspace{0.3cm}
{\em $^{\rm b}$Department of Theoretical Physics,
Lebedev Physical Institute, \\
 117924 Moscow, Russia\\}
\vspace{0.3cm}
{\em $^{\rm c}$Theory Division, CERN, CH-1211 Geneva 23,
Switzerland\\}
\vspace{0.3cm}
{\em $^{\rm d}$Department of Physics,
P.O.Box 9, 00014 University of Helsinki, Finland\\}

\vspace{0.7cm}
{\bf Abstract}

\end{centering}

\vspace{0.3cm}\noindent
We study the universality class and critical properties of the
electroweak theory at finite temperature. Such critical behaviour 
is found near the endpoint $m_H=m_{H,c}$ of the line of first order
electroweak phase transitions in a wide class of theories, including
the Standard Model (SM) and a part of the parameter space of the
Minimal Sypersymmetric Standard Model (MSSM).  We find that the
location of the endpoint corresponds to the Higgs mass $m_{H,c}=
72(2)$ GeV in the SM with $\sin^2\theta_W=0$, and $m_{H,c}<80$ GeV
with $\sin^2\theta_W=0.23$. As experimentally $m_H>88$ GeV, there is no
electroweak phase transition in the SM. We compute the corresponding
critical indices and provide strong evidence that the phase
transitions near the endpoint fall into the three dimensional Ising
universality class.

\vfill
\noindent
CERN-TH/98-08\\
NORDITA-98/30HE\\
May 1998

\vfill

\end{titlepage}

\section{Introduction}
The finite temperature phase transition in the Standard Model (SM) is
known to be a first order transition for small and a crossover for
large Higgs masses \cite{isthere}. In between there is a critical
region at about $m_H=$ 75 GeV \cite{karschnpr,gurtler1} (see also~\cite{bph}). 
The purpose of this paper is to study this critical region in a detailed
manner. We shall show that the universality class of the endpoint is
that of the three-dimensional (3d) Ising model. We also obtain the
value $m_{H,c}<80$ GeV for the endpoint Higgs mass in the SM. Given
that the experimental 95\% C.L. lower limit is $m_H> 87.9$ GeV
\cite{mHlower}, there would be no phase transition, only a crossover,
if the physics were that of the SM.

Universality implies a tremendous simplification in the degrees of
freedom of the system. Here the first step is the removal of all
fermionic and all non-static (not constant in imaginary time) bosonic  
fields~\cite{dimred}. Equivalently, one integrates out all             
fields with masses $\gsim\pi T$. This works for equilibrium phenomena
in the high $T$ small coupling limit. 
Furthermore, all masses $\sim gT$
can also be integrated out. Hereby one obtains a 3d effective theory
$S[B_i,A_i^a,\phi_k],\,\,i,a=1,2,3,\,\, k=1,\ldots,4$, with  
SU(2)$\times$U(1) symmetry and a fundamental doublet $\phi$ \cite{generic}. 
The superrenormalizable 3d theory provides a very good approximation to
high $T$ 4d physics. The accuracy of the effective description has been
discussed in detail in \cite{generic}; further corrections to
the effective action can also be computed.

The previous steps can be performed perturbatively, but further
progress is only possible with numerical lattice Monte Carlo
techniques (for reviews, see \cite{jansen,rummureview}).  In terms of
SM physics, these show the existence of a line of first order phase
transitions $T_c = T_c(m_H)$, $m_H<m_{H,c}$, which ends at
$T_c(m_{H,c})$ and turns into a crossover at $m_H>m_{H,c}$. When
approaching the endpoint along the first order line, the mass of one
of the scalar excitations seems to go to zero suggesting
\cite{nonpert} that ultimately all other masses could be integrated
out, leaving near $(m_{H,c},T_c(m_{H,c}))$ a final effective theory
$S_\rmi{crit}[\phi']$ containing only one scalar degree of freedom $\phi'$.

To be more precise, the electroweak theory with a Higgs doublet
contains a massless vector excitation, corresponding to the
hypercharge field high in the symmetric phase and to the photon deep
in the Higgs phase. The fact that this state is massless at any
temperature ensures the ``topological'' similarity of the phase
diagrams of the SU(2)+Higgs and SU(2)$\times$U(1)+Higgs
theories~\cite{su2u1}. Moreover, the lowest order gauge-invariant
coupling of a real scalar to a vector field $\phi'
F_{ij}F_{ij}$ has a dimensionality greater than 3, and thus the scalar
is decoupled from the massless vector in the infrared.  Hence, for
discussing the universal properties of the theory near the endpoint,
we can work with the SU(2)+Higgs theory $S[A_i^a,\phi_k]$,
disregarding the U(1) interactions. We shall in the following show
that the endpoint of this theory belongs to the 3d Ising universality
class~\cite{3dising,HaPinn}. The universality class of the 3d O(4)
invariant spin model~\cite{kanaya,toussaint}, which has also been
proposed as a possible candidate \cite{karsch2}, can be ruled out.

The matching of the {\em continuum\,} theories 
$S[A_i^a,\phi_k]\to S_\rmi{crit}[\phi']$ is a delicate
issue and at this stage we do not
determine the couplings of $S_\rmi{crit}$, only its universality
class.  We first discretize the continuum theory at fixed lattice
spacing $a$ and determine the critical properties of the discretized
theory near the endpoint $(m_{H,c},T_c(m_{H,c}))$.  This is done by
studying the properties of probability distributions of various
observables (hopping term, $(\phi^2-1)^2$, etc) averaged over a
finite-volume system (we consider the theory in a cubic box with
periodic boundary conditions). We obtain the joint probability
distribution of up to 6 observables and analyze it in two ways:

1. We compute the fluctuation matrix, study the dependence of its
eigenvalues on the volume and obtain critical indices, which turn out
to be consistent with those of the 3d Ising model,

2. We show that for a certain pair of observables, which may be denoted
as $M$-like 
and $E$-like, 
the joint probability distribution has a very nontrivial form which matches
closely the joint distribution of magnetization $M$ and energy $E$ of
the 3d Ising model in a box of the same geometry. This guarantees that
not only the critical indices, but also higher moments agree with those
of the 3d Ising model.
As a byproduct we obtain the mapping of the 6-dimensional operator
space to the Ising model.  This is a fixed lattice spacing
version of the critical mapping: $S_6[A^a_i,\phi_k,a]\to   
S_{\rm Ising}$.

Our method has much in common with, and can be considered as the
generalization of, the method used by Alonso {\em et al} \cite{alonso}
to locate and study the endpoint of the first order transition line
separating the Higgs and confinement phases in the 4d U(1)+Higgs model,
and the method developed by Bruce and Wilding \cite{wilding} for the
study of the liquid-gas critical point, both of which rely on
considering two-dimensional probability distributions and finding
the $M$-like and $E$-like directions. However, our method as well as
some aspects of the critical behaviour of our system, differ in many
important respects from those in~\cite{alonso,wilding}.

Finally, an extrapolation to $a\to0$ will have to be made. There is no
change in the universal properties. However, this extrapolation is
needed to get the continuum value of the (non-universal) quantity
$m_{H,c}$. As mentioned above, this in conjunction with the
experimental lower limit implies that the critical region of the 3d
effective theory can only be physically relevant in a beyond-the-SM
electroweak theory, such as the MSSM.

The plan of the paper is the following. We formulate the problem in
Sec.~\ref{sec:formulation} in some more detail, and outline its
solution in Sec.~\ref{sec:outline}.  In Sec.~\ref{sec:spin} we review
the basic properties of O(N) spin models. Sec.~\ref{sec:critical}
contains a detailed presentation of the method of determining the
universality class of the 3d SU(2)+Higgs theory. Asymmetry effects are
studied in Sec.~\ref{sec:asymm}.  In Sec.~\ref{sec:summary} we
summarize the results for the critical properties and for $m_{H,c}$,
and we conclude in Sec.~\ref{sec:conclusions}.

\section{Formulation of the problem}\la{sec:formulation}

At finite temperatures, the static bosonic correlators in the Standard
Model and many of its extensions can be derived from the 3d effective
action (as discussed above, we omit the U(1) interactions)
\ba
S & = & \int\! d^3x \biggl[
\frac{1}{2}\tr F_{ij}F_{ij}+
(D_i\phi)^{\dagger}(D_i\phi)+
m_3^2\phi^{\dagger}\phi+
\lambda_3 (\phi^{\dagger}\phi)^2\biggr],
\label{action}
\ea
in standard notation. This is a continuum field theory characterized
by the dimensionful gauge coupling $g_3^2$ and by the dimensionless
ratios
\be
x=\lambda_3/g_3^2, \quad
y=m_3^2(g_3^2)/g_3^4, \label{xy}
\ee
where $m_3^2(\mu)$ is the renormalized mass parameter in the $\msbar$
scheme. The relations of $g_3^2,x,y$ to the full theory are computable
in perturbation theory, and the relative accuracy thus obtained for
non-vanishing one-particle irreducible Green's functions $G$ that
conserve parity, C and CP is~\cite{generic}
\be
\frac{\delta G}{G}
\lsim O(g^3), \la{errors}
\ee
where $\delta G$ is the error in $G$. The fact that there is a
suppression of error arises from the ratio of the scales left and integrated
out, ${\cal O}(gT/T)$, ${\cal O}(g^2T/gT)$, and the third power from
the types of higher order operators that have been
neglected~\cite{generic}. Hence, for small coupling, the physics of the
4d theory can be described accurately with a much simpler 3d theory.
Explicit derivations have been given in~\cite{generic,mssm}.

Instead of using the $\msbar$ scheme, 
the 3d continuum theory of \eq\nr{action}
can as well be regulated by using a lattice with the lattice constant
$a$. The action then is
\begin{eqnarray}
S&=& \beta_G \sum_\bfx \sum_{i<j}(1-\fr12 \tr P_{ij}) \nonumber \\
 &-& \beta_H \sum_\bfx \sum_i
\fr12\tr\Phi^\dagger(\bfx)U_i(\bfx)\Phi(\bfx+i)
\nonumber \\
 &+& \sum_\bfx
\fr12\tr\Phi^\dagger(\bfx)\Phi(\bfx) + \beta_R\sum_\bfx
 \bigl[ \fr12\tr\Phi^\dagger(\bfx)\Phi(\bfx)-1 \bigr]^2
\la{latticeaction} \\
&\equiv& S_G+S_\rmi{hopping}+S_{\phi^2}+S_{(\phi^2-1)^2},
\nonumber
\end{eqnarray}
in standard notation [$\Phi$ is the matrix $\Phi=(i\sigma_2\phi^*,
\phi)]$. The two actions in \eqs\nr{action}, \nr{latticeaction} give
the same physics in the continuum limit $a \rightarrow 0$ if the three
dimensionless parameters $\beta_G, \beta_H,\beta_R$ in
\eq\nr{latticeaction} are related to the three dimensionless
parameters $g_3^2a,x,y$ in \eq\nr{action} by the following equations
\cite{contlatt}:
\begin{eqnarray}
\beta_G & = & {4\over g_3^2a}, \la{betag}\\
\beta_R & = & {\beta_H^2\over\beta_G}x, \la{betar}\\
y & = &
{\beta_G^2\over8}\biggl({1\over\beta_H}-3-
{2x\beta_H\over\beta_G}\biggr)+{3\Sigma\beta_G\over32\pi}
(1+4x)
\nonumber\\
&&+{1\over16\pi^2}\biggl[\biggl({51\over16}+9x-12x^2\biggr)
\biggl(\ln{3\beta_G\over2}+\zeta\biggr)+
4.9941 +5.2153 x\biggr],
\la{y}
\end{eqnarray}
where $\Sigma=3.1759115$ and $\zeta=0.08849(1)$.
The two numbers 4.9941 and 5.2153 are specific for the SU(2)+Higgs theory.

The approach to the continuum limit can be accelerated by removing the
$O(a)$ errors analytically~\cite{moore}. For $g_3^2,x$, this can be
achieved by reinterpreting the simulation results employing
\eqs\nr{betag}, \nr{betar} as corresponding to
\ba
\frac{4}{(g_3^2)_{\rm improved}a} & =  &
{4\over g_3^2a} - 0.6674, \\
x_{\rm improved} & = & x - \frac{1}{\beta_G}
(0.018246 + 0.195709 x + 0.583880 x^2). \la{eq:moore}
\ea
For $y$ the issue is more involved, see~\cite{moore}.

\begin{figure}[t]

\centerline{\hspace{-3.3mm}
\epsfxsize=9cm\epsfbox{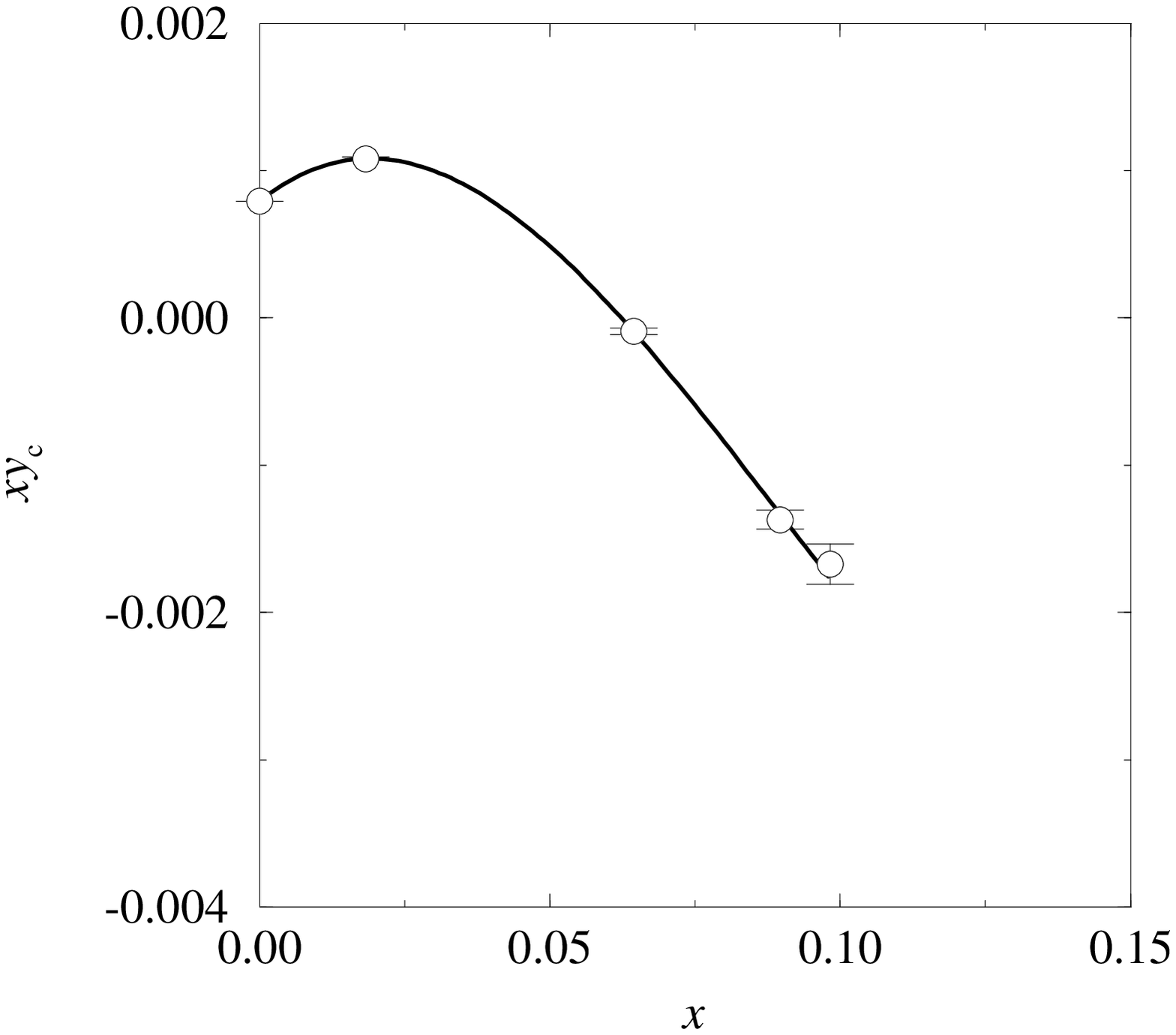}
\hspace{-1.5cm}
\epsfxsize=9cm\epsfbox{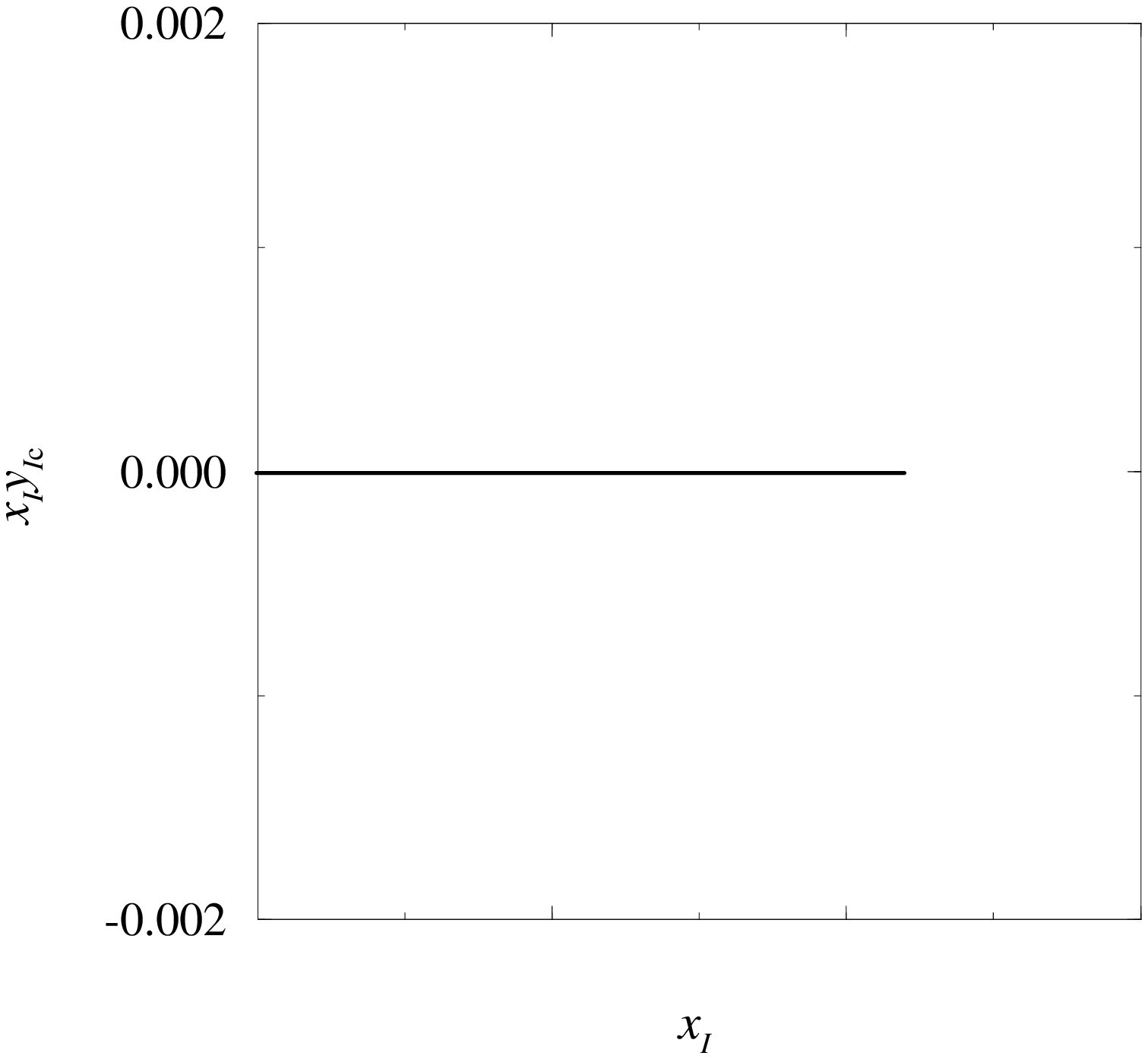}}

\vspace*{-4.5cm}

\caption[a]{{\em Left:}
The phase diagram of the 3d SU(2)+Higgs theory. The datapoints are  
from~\cite{nonpert} and from this paper. The value of $x\cdot y_c$ at
$x\to 0$ is given by the 1-loop effective potential and is hence known
analytically.
{\em Right:}
The phase diagram of the 3d scalar $\phi^4$ theory
in \eq\nr{isingaction}. 
The value of
$x_{I}$ at the endpoint has to our knowledge not been determined.
The number 0.002 on the vertical axis is symbolic, 
as the figure is scale invariant in this direction.}
\la{fig:xyc}
\end{figure}

The phase structure of the theory in \eq\nr{action} is shown in
Fig.~\ref{fig:xyc}. There is a first order line $y=y_c(x)$ for $x<x_c$;
for $x>x_c$ there is only a crossover. The first order
line is localised by using the lattice action in \eq\nr{latticeaction}
at some finite lattice spacing $a$ and system volume $V$, finding a
two-peak distribution in the measurements of any gauge invariant
observables and performing the limits $V\to\infty$ and $a\to0$. At
small $x\ll x_c$ the two peaks are very asymmetric and well separated,
signalling a strong transition. When $x$ approaches $x_c$, the two
peaks become more symmetric and approach each other. One of the masses,
$\mH$ measured by the correlator of $\Phi^\dagger\Phi$, becomes smaller
than the other masses. One expects that $\mH\to0$ at $x_c$ so that the
transition is of the second order there.

In practice, one takes the lattice theory in \eq\nr{latticeaction} at some
fixed $\beta_G$ and finds the location of the first order line in the
plane of the remaining two parameters $(\beta_H, \beta_R)$. To approach
the continuum limit, one has to repeat the study at higher values of
$\beta_G$ (we have used $\beta_G = 5$, 8 and 12).

Since the mass $\mH$ is expected to vanish at the endpoint $x_c$, a
natural question arising is whether the effective theory in
\eq\nr{action} could be further reduced leaving only the lightest
excitations in the final action~\cite{patkos,karsch1,nonpert}. Knowing
the effective theory would also imply knowledge of the universality
class of the second order transition at~$x_c$. Unfortunately, no
systematic perturbative derivation of such an effective action has been
found. One problem is that in perturbation theory the vector
excitations are massless in the symmetric phase, leading to
IR-problems, whereas non-perturbatively they are massive and are to be
integrated out. Thus the construction of the effective theory has to be
non-perturbative.

As to the functional form of the effective theory, the fact that there
is only one light physical scalar degree of freedom, the Higgs
particle, naturally leads to the suggestion \cite{isthere,nonpert} that
the corresponding $\msbar$ continuum effective field theory is
\be
S = \int\! d^3x \biggl[
\fr12 \partial_i \phi\partial_i \phi+
h_I \phi + \fr12 m_I^2 \phi^2 + \fr14 \lambda_I \phi^4 \biggr].
\la{isingaction}
\ee
We discuss the renormalisation and discretization of this theory in the
Appendix. The theory in \eq\nr{isingaction} is characterized by the
scale $\lambda_I$ and by the dimensionless ratios
$y_I=h_I/\lambda_I^{5/2}$, $x_I=m_I^2(\lambda_I)/\lambda_I^2$, where
$m_I^2(\mu)=-6\lambda_I^2/(16\pi^2) \log(\Lambda_m/\mu)$ is the running
mass parameter in the $\msbar$ scheme, 
and $\Lambda_m$ is scale invariant. An otherwise possible cubic term
can always be shifted away and this makes $h_I$ scale independent.
Higher order operators could also exist, but they give contributions
suppressed by ${\cal O}(\mH/m_W)$ where $m_W$ stands for all masses,
like the inverse vector correlation length, which remain finite
at~$x_c$.

In the critical region the theory in \eq\nr{isingaction} is in the
same universality class as the 3d Ising model in an
external magnetic field $h$,
\be
Z=\sum_{\{s_i\}}\exp[\beta\sum_{\langle ij\rangle} s_is_j +
h\sum_i s_i],   \qquad s_i = \pm 1.
\la{ising}
\ee
Here $\beta$ is the inverse temperature, the spins $s_i$ are located    
at the sites of a simple cubic lattice, and $\langle ij\rangle$ denotes 
the pairs of nearest neighbours.
\smallskip

To be more specific about constructing an effective theory such as the
one in \eq\nr{isingaction} (or, for a finite lattice spacing, 
the one in \eq\nr{ising}), 
there are two questions to be considered:

1. Which is the functional form (the degrees of freedom;
   the symmetries; the universality class)
   of the effective theory?

2. Which is the mapping between the parameters $g_3^2,x,y$ of the
   original theory, and those of the effective theory (such as
   $\lambda_I,x_I,y_I$ of the scalar theory)?

The latter of the two questions is much more difficult than the former
one and its solution will not be attempted here. The reason for the
difficulty is that the mapping is non-universal and depends on the
detailed UV-properties of the original theory. In particular, since the
mapping has to be done non-perturbatively on the lattice,

(a) there are finite lattice spacing effects in the lattice formulation
in \eq\nr{latticeaction} of the original theory in \eq\nr{action}.
These should be identified and removed by an extrapolation to the
continuum limit.

(b) there are finite lattice spacing effects in the lattice formulation in
\eq\nr{scalarlattaction} of the effective theory in
\eq\nr{isingaction}. These should be controlled in a similar way.

(c) apart from the lattice spacing effects, there are also higher order
operators in the effective theory in \eq\nr{isingaction}
applying in the continuum limit, due
to the degrees of freedom which have been integrated out. The effects
of these are suppressed by $O(\mH/m_W)$, but they induce errors
immediately when one goes away from the critical point
(or is at a finite volume).

For the determination of the universality class, in contrast, none of
these problems arises. By definition, the universal properties are
insensitive to the UV. Hence no extrapolation is needed to overcome
(a), and a single finite lattice spacing may be used. However, there is
a price to be paid for a finite lattice spacing, which is that there
are infinitely many gauge-invariant operators available, and to find
the optimal projections on the critical directions, one should take as
large an operator basis as possible. No extrapolation to $a\to 0$ is
needed for (b), either, and one can directly compare with the known
properties of the spin models in the same universality class as the
continuum theory considered. Finally, the non-universal errors
responsible for (c) vanish at the critical point (but may induce
corrections to scaling, etc). In the following we
shall mostly concentrate on the universal properties.

\section{Outline of the solution of the problem}\la{sec:outline}

To study whether the universality properties of the theory in
\eq\nr{action} match those of the scalar theory in
\eq\nr{isingaction}, it is sufficient to compare the critical
properties of the lattice SU(2)+Higgs theory in \eq\nr{latticeaction}
directly with those of the Ising model, \eq\nr{ising}.

To gain insight into what is happening at the critical point,
it is very helpful to consider two-dimensional probability
distributions (joint distributions of two observables)
\cite{alonso,wilding}. The motivation is as follows. In the theory
in \eq\nr{latticeaction} we have a two-dimensional parameter plane where we
find a first order phase transition line which ends in a critical point.
In this respect, our system is very similar to such well known systems
as the liquid-gas phase transition (where the parameters are
the temperature and the pressure) and the Ising model in an external magnetic
field (the parameters being the temperature $1/\beta$ and the field $h$).

It has already been checked to considerable precision, both
experimentally and by Monte Carlo simulations \cite{wilding}, that in
the case of the liquid-gas transition, not only the topology of the critical
point is similar to that of the Ising model in an external field, but also
the universality class is the same: one can find a linear
mapping of a small area around the liquid-gas critical point to the
corresponding area around the Ising critical point, such that both
systems behave in exactly the same way (up to corrections to scaling
at a finite volume).

Our aim is to provide the evidence that the same is also true for our
system. Thus we expect to find at the endpoint a temperature-like
($t$-like) direction in the parameter space $(\beta_H, \beta_R)$, going
tangentially to the phase transition line, and a magnetic field-like
($h$-like) direction, corresponding to the $t$ and $h$ directions
of the Ising model.

Our system, as well as the liquid-gas system, lacks the exact symmetry
$h \to -h$, which is characteristic of the Ising model. As has been
shown for the case of the liquid-gas system, in this case the $h$-like
and $t$-like directions are not necessarily orthogonal
in the $(\beta_H,\beta_R)$ space
\cite{MerminRehr,wilding}. Thus, in the vicinity of the critical point
of our system, the orthogonal (in the sense $\langle\Delta E\Delta M
\rangle=0$)
energy-like and magnetization-like observables $\tE$ and $\tM$, 
being derivatives of the free energy over
the $t$-like and $h$-like directions in the $(\beta_H, \beta_R)$ space,
are going to be certain linear combinations of the corresponding terms
in the action, that is, of $S_\rmi{hopping}$ and $S_{(\phi^2-1)^2}$, 
with a possibly non-orthogonal proportionality matrix. In
the Ising model the observables $\tE$ and $\tM$ just correspond
to the first and second terms in \eq\nr{ising}.

Before attempting to study the 
probability distributions of $\tE$ and $\tM$, 
one has to determine the coefficients of these
linear combinations. Thus one arrives at the idea of looking at {\em  
two-dimensional} probability distributions: for every configuration
generated by Monte Carlo one computes and stores two numbers,
$S_\rmi{hopping}$ and $S_{(\phi^2-1)^2}$, thus obtaining their joint
probability distribution.

Such distributions are very useful, as they contain a lot
of information. Having collected this distribution at some point in the
parameter space close to the critical point, one can later
find the $E$-like and $M$-like directions;
compute the 1-dimensional probability distribution for any
linear combination (or even arbitrary function) of
$S_\rmi{hopping}$ and $S_{(\phi^2-1)^2}$; 
refine the estimate for the position of the critical point;
and reweight the data to the more precisely determined critical point.


\begin{figure}[t]

\centerline{
\epsfxsize=7cm \epsffile{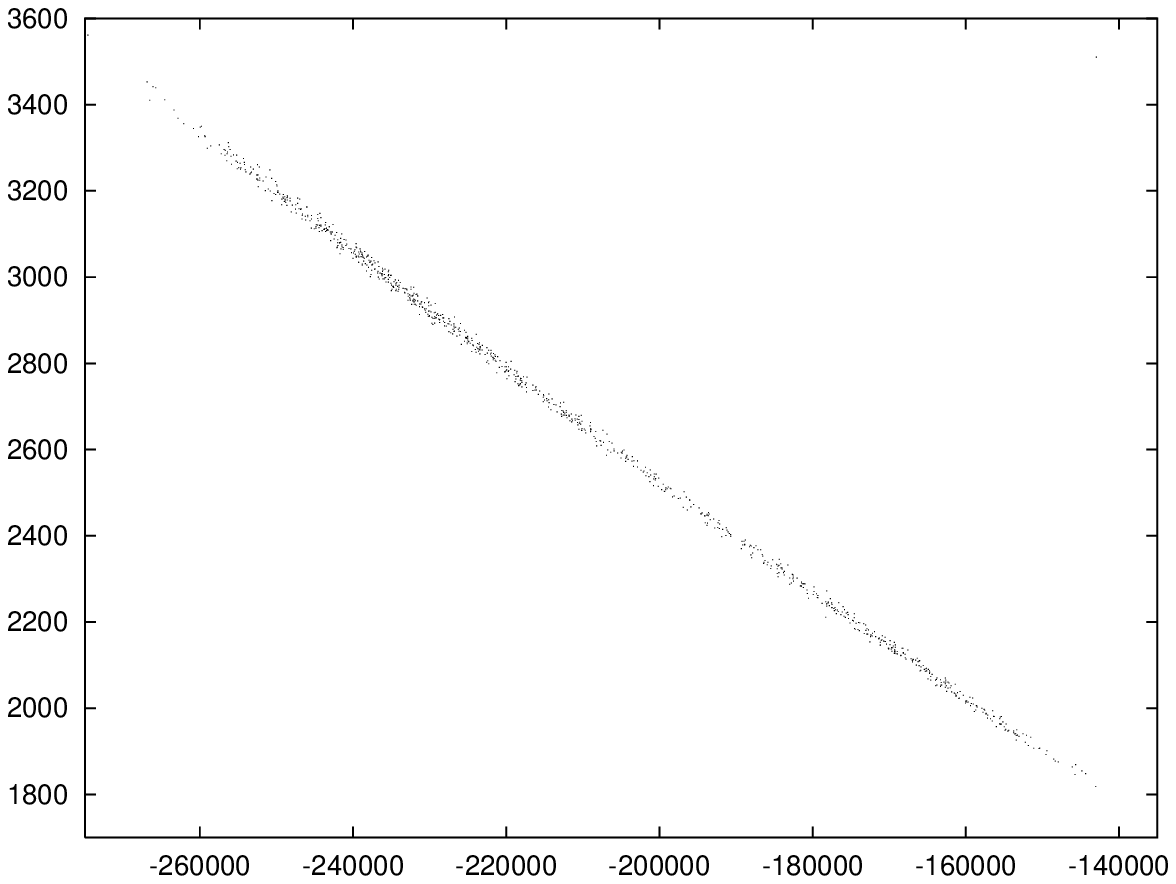}
\epsfxsize=7cm \epsffile{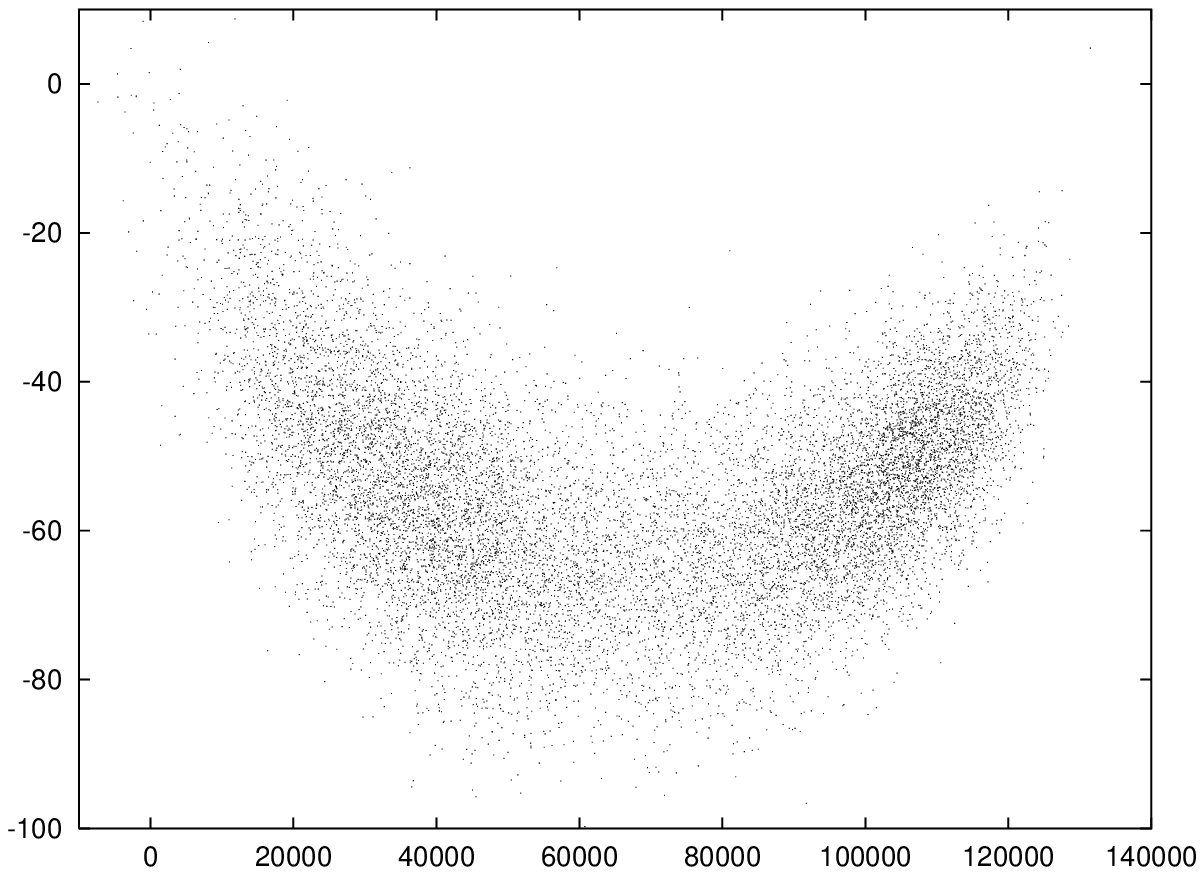}}

\vspace*{0.3cm}

\centerline{(a) \hspace*{7cm} (b)}

\vspace*{0.3cm}

\centerline{
\epsfxsize=7cm \epsffile{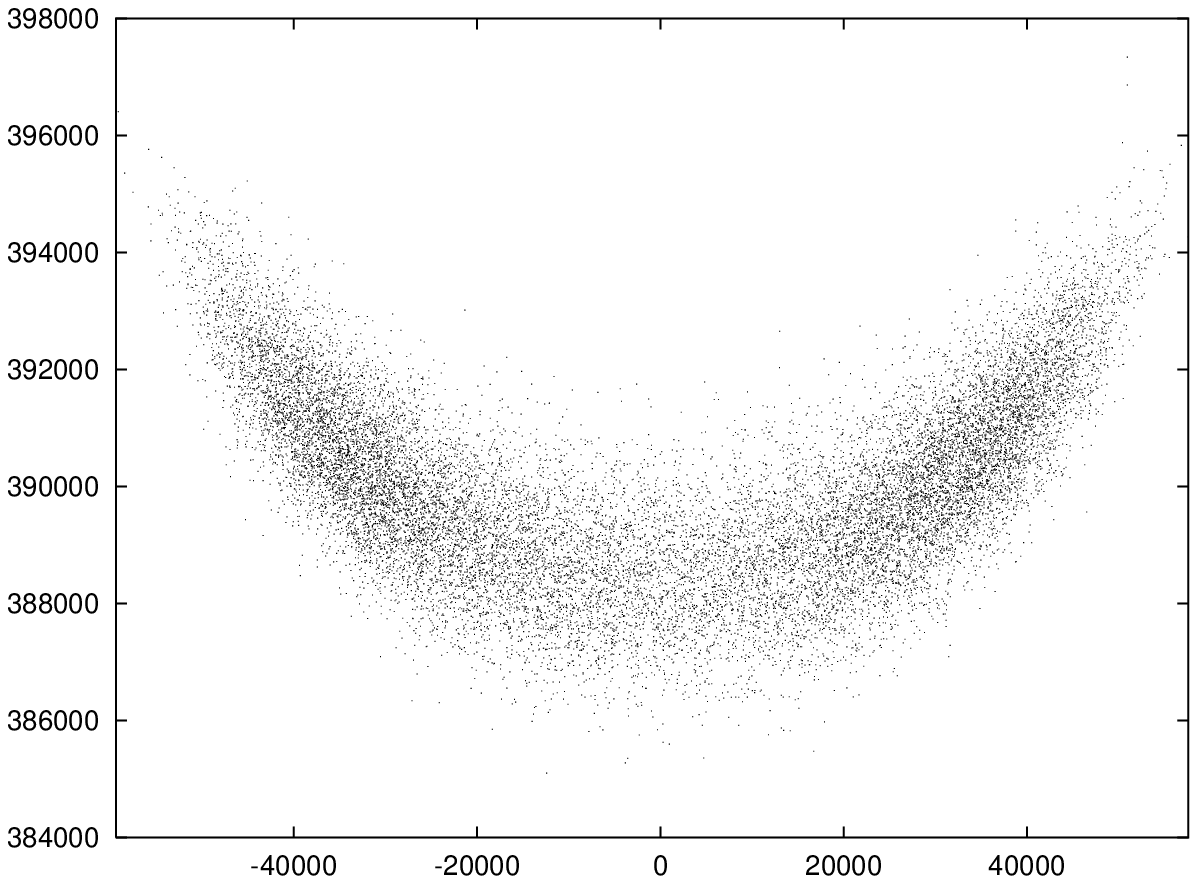}}

\vspace*{0.3cm}

\centerline{(c)}

\vspace*{0.3cm}

\caption[a]{
(a) 1000 configurations from the Monte Carlo simulation
of the theory in \eq\nr{latticeaction}, represented by points in the
$S_{(\phi^2-1)^2}$ vs.\ $S_\rmi{hopping}$ plane, for $x=0.105253,
\beta_G=8,\beta_H=0.349853, V=48^3$.
\la{352-1000}
(b) 13822 configurations of the same system, for the same parameter
values, after a shift and rotation in the
coordinate plane. The angle of rotation is chosen to make
the elongated distribution in (a) go approximately
horizontally.
\la{352t-all}
(c) 20000 configurations of the 3d Ising model
on a $58^3$ lattice at
the critical point $\beta_c=0.221654,\ h=0$,
on ``minus the energy'' ($0<\sum_{\langle ij\rangle}\delta_{s_is_j}<3\cdot
58^3$)
vs. magnetization\ ($-58^3<\sum_i s_i<58^3$) plane.
\la{p51}}
\end{figure}


To give a concrete view of what is happening, Fig.~\ref{352-1000}(a)
shows the distribution of 1000 configurations obtained near the
endpoint, plotted on the $S_{(\phi^2-1)^2}$ (vertical axis) vs.\
$S_\rmi{hopping}$ (horizontal axis) plane. One can see that the
distribution appears to be extremely elongated, and the points tend to
concentrate on its sides. The density in the middle is somewhat
smaller, thus a two-peak distribution is produced by projecting on
either of the axes. Rotating the coordinate system in such a way that
the new $x$-axis goes along our distribution, while the new $y$-axis is
orthogonal to it, and changing scales, 
we obtain the distribution plotted in
Fig.~\ref{352t-all}(b). This should be compared with Fig.~\ref{p51}(c)
which shows 20000 configurations of the 3d Ising model at the critical
point on ``minus the energy'' (vertical axis: $\sum_{\langle
ij\rangle}\delta_{s_is_j}$) vs. magnetization (horizontal axis: $\sum_i
s_i$) plane. The qualitative similarity with Fig.~\ref{352t-all}(b) is
striking. At the same time, there are small discrepancies: the
distribution in Fig.~\ref{352t-all}(b) is slightly asymmetric, and
considerably thicker 
than the one in
Fig.~\ref{352t-all}(c). Comparing with Fig.~\ref{352t-all}(c), one
observes that the new $x$- and $y$-axes correspond, with reasonable
precision, to $M$-like and $E$-like directions. Note that a projection
onto the $M$-like direction produces a two-peak probability
distribution, while the $E$-like projection is single-peaked.

The task now is to put the similarity between SU(2)+Higgs and Ising
models on a quantitative basis, and to demonstrate that the
discrepancies disappear when $V\to\infty$. Note, however, that already
the distribution in Fig.~\ref{352t-all}(b) suggests that O(4)
universality is excluded; O(4) would not produce 2-peak distributions
like those in Fig.~\ref{p51}(c) and thus cannot match the data in
Fig.~\ref{352t-all}(b) [see Fig.~\ref{spinmodels}(c)].

At this point it is interesting to compare our two-dimensional
distributions, depicted in Fig.~\ref{352t-all}, with those obtained in
\cite{alonso} near the endpoint of the first order phase transition
line separating the Higgs and confinement phases in the 4d U(1)+Higgs
theory. First, the distributions in \cite{alonso} demonstrate almost
independent fluctuations of $\tM$ and $\tE$, while in our
case the ``boomerang'' shape implies that their fluctuations are
obviously {\em not\/} independent. Secondly, the distributions in
\cite{alonso} do not show any visible asymmetry, while we see a clear
residual asymmetry that can be attributed to corrections to scaling.
These differences are probably due to the fact that while the critical
point of the 4d model corresponds to a trivial effective theory (this is
corroborated by the critical indices obtained in \cite{alonso}, which
are compatible with the mean field values), in our case the effective
theory, being 3-dimensional, is governed by a nontrivial fixed point.

\section{Spin models}\la{sec:spin}

Let us start by reviewing some basic properties of simple spin models.
The Ising model is defined by \eq\nr{ising}. For O(N) models, on the
other hand, the spins $s_i$ of the Ising model are replaced by
$N$-dimensional unit vectors ${\bf s}_i$, and the partition function
becomes
\be
Z=\int\{ d {\bf s}_i\}\exp[\beta\sum_{\langle ij\rangle} {\bf s}_i
\cdot {\bf s}_j +
{\bf h} \cdot \sum_i {\bf s}_i].
\la{O(N)}
\ee
Let us call the first term in the exponent the energy variable $E$,
and the next term the magnetic variable $M$:
\be
E= - \sum_{\langle ij\rangle} {\bf s}_i
\cdot {\bf s}_j, \quad
M=\frac{\bf{h}}{|\bf{h}|} \cdot \sum_i {\bf s}_i.
\ee
The characteristics of the model at the critical couplings
$\beta=\beta_c$, $|{\bf h}|=|{\bf h}_c|=0$ are contained in probability
distributions in the $(M,E)$-plane, as a function of the volume of the
system. When the variances of the distributions are normalized to
unity, the distributions have a universal form in the large volume
limit; the results for the Ising, O(2) and O(4) models are shown in
Fig.~\ref{spinmodels}. To reduce statistical noise, these contour
plots, as well as the contour plots in the following pictures, have
been smoothed by $3 \times 3$ matrix averaging. That is, before
plotting, the occupation number in every bin is replaced by the average
over 9 bins forming a square around it. The bin size has been chosen
sufficiently small so that smoothing does not induce any significant
broadening of the peaks.

\begin{figure}[p]

\centerline{
\epsfysize=6.2cm \epsfbox[36 40 539 468]{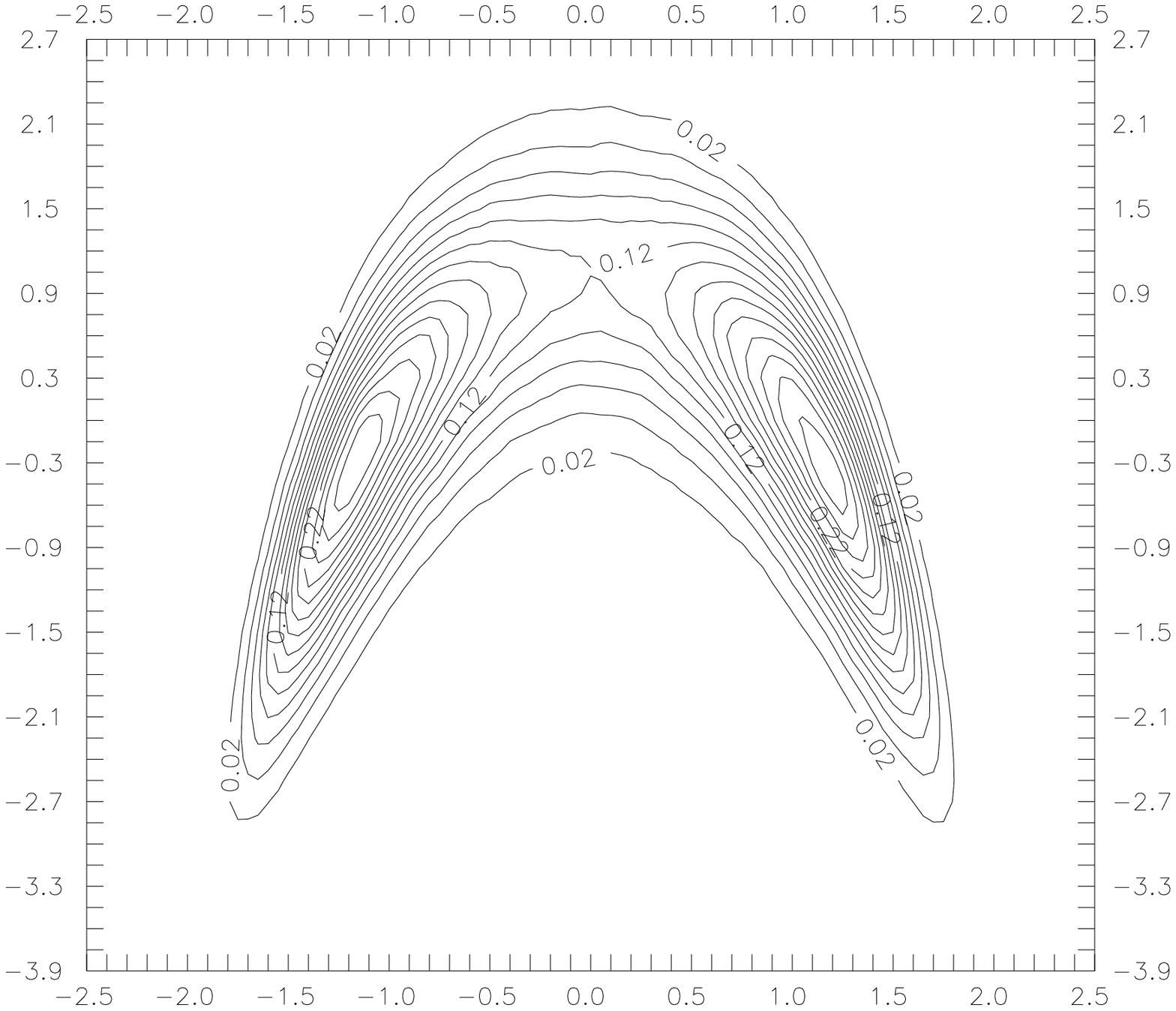}
\hspace*{5mm}
\epsfysize=6.2cm \epsfbox[36 40 539 468]{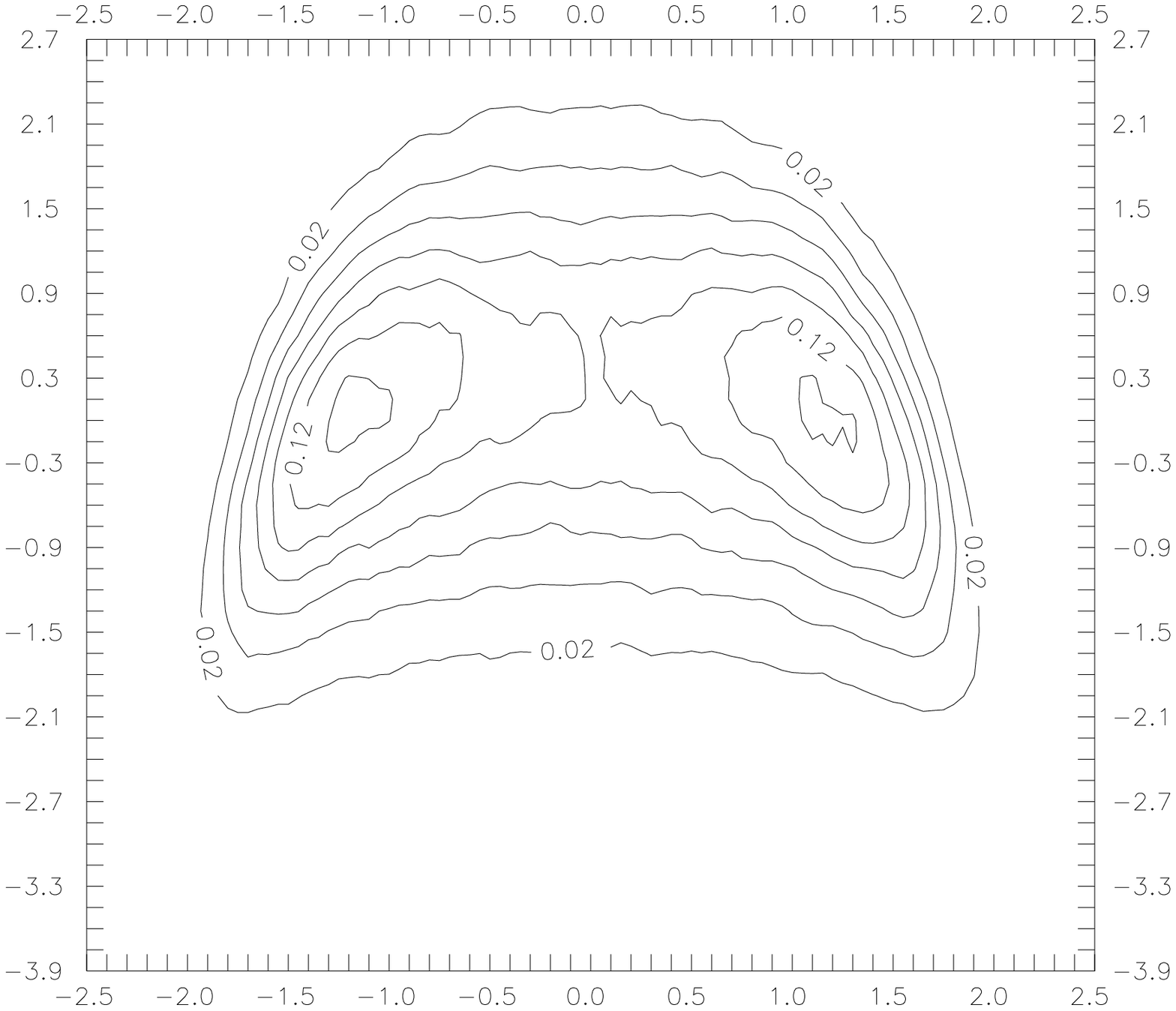}
}
\centerline{(a) \hspace*{7cm} (b)}
\vspace*{5mm}
\centerline{
\epsfysize=6.2cm \epsfbox[36 40 539 468]{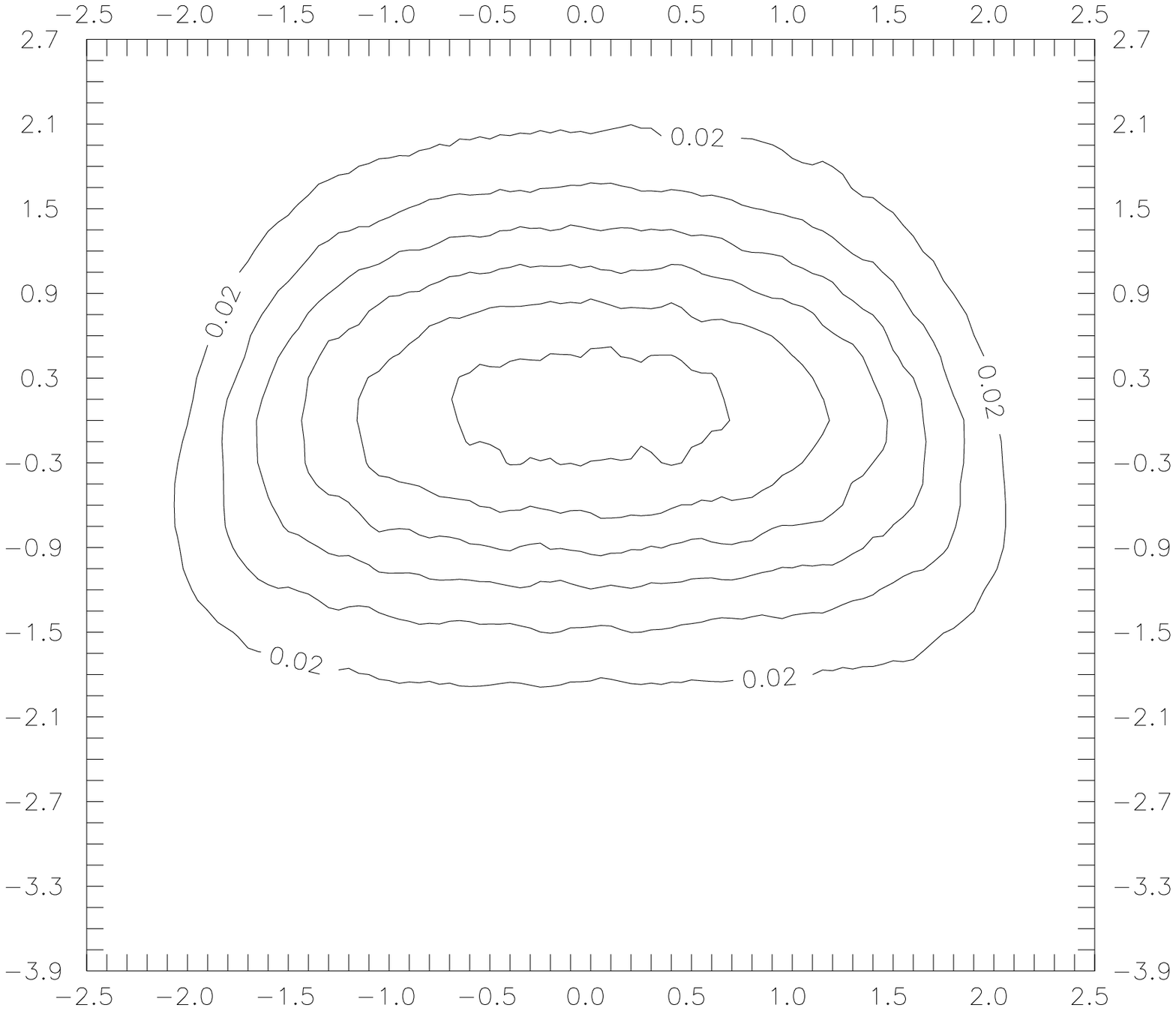}
}
\centerline{(c)}


\caption[a]{
The smoothed and normalized probability distributions,
at the critical point, for
(a) the Ising model at the volume $58^3$,
(b) the O(2) spin model at $64^3$,
(c) the O(4) spin model at $64^3$.
The $x$-axis is the magnetic direction and
the $y$-axis the energy direction.}
\la{spinmodels}
\end{figure}

To quantify the characteristics of the probability distributions, one
can compute different moments. Let $\Delta E \equiv E-\langle
E\rangle$, $\Delta M \equiv M - \langle M\rangle$. Then the moments of
interest are:

1. Second moments (specific heat, magnetic susceptibility):
\ba
\chi_E & = & \langle (\Delta E)^2 \rangle/L^3, \label{chi_E}\\
\chi_M & = & \langle (\Delta M)^2 \rangle/L^3, \label{chi_M}
\ea
where $L^3$ is the volume of the system. The behaviour of
these moments as a function of $L$ is characterized by critical
exponents:
\be
\chi_E \propto L^{\alpha/\nu}, \quad
\chi_M \propto L^{\gamma/\nu}. \label{gamma}
\ee
The known results for the exponents appearing here
are~\cite{ZJ,kanaya,O4refs,gz,hpv}:
\smallskip

\centerline{
\begin{tabular}{|c|c|c|c|c|c|}
\hline
model & $\gamma$ & $\alpha$ & $\nu$ & $\gamma/\nu$ & $\alpha/\nu$ \\
\hline
Ising & 1.24 &  0.11  & 0.63 & 1.96 &  0.17  \\
O(2)  & 1.32 & -0.01  & 0.67 & 1.96 & -0.015 \\
O(4)  & 1.47 & -0.25  & 0.75 & 1.96 & -0.33  \\
\hline
\end{tabular}}
\smallskip
\smallskip

\noindent
The exponent $\nu$ is the correlation length critical exponent.

2. Higher moments. These characterize for instance the symmetry
features of the probability distributions. In particular, all spin
models have, for $n=1,2,\ldots$,
\be
\langle (\Delta M)^{2n+1}\rangle =0.
\ee
As examples of non-zero values, let us mention that for the 3d Ising
model in a large cubic box with periodic boundary conditions, the value
of the following ratio is known with high precision \cite{Bloete}:
\be
\frac{\langle (\Delta M)^4\rangle}{\langle (\Delta M)^2\rangle^2}
 = 1.604(1). \label{M^4}
\ee
The asymmetry of the energy distribution of the same system is
characterized by
\be
\frac{\langle (\Delta E)^3\rangle}{\langle (\Delta E)^2\rangle^{3/2}}
 \approx -0.36 \label{E^3}\\
\ee
(this value corresponds to the simple cubic Ising model on a $58^3$
lattice, where deviations from scaling due, in particular, to the
presence of a large regular part in the energy, are still
non-negligible).

These ratios, as well as the critical exponents, are universal
quantities which can be used to quantify the similarity or
dissimilarity of the endpoint of the SU(2)+Higgs theory with different
spin models.

\section{Detailed study of the critical region}\la{sec:critical}

As discussed above, 
our computational strategy is based on collecting joint probability
distributions of several observables (initially two, as in
Fig.~\ref{352t-all}, and then up to six, as discussed below)
for the system in \eq\nr{latticeaction} in a cubic box with periodic
boundary conditions. These are used to

 1. find the position of the critical point,

 2. determine the $M$-like and $E$-like directions in the
    space of observables,

 3. perform finite size scaling (FSS) to compute critical indices,
    by studying how the fluctuations of $M$-like and $E$-like
    observables 
    at the critical point
    depend on the system size,

 4. determine higher moments, such as the skewness of $E$.

\vspace{4mm} Our analysis is based on simulations at $\beta_G = 5$
with the lattice sizes and statistics shown in Table~\ref{tab:stat}.
In each case the measurements are separated by 4 overrelaxation sweeps
and one heat bath/Metropolis sweep. The update algorithms used are
described in Ref.~\cite{nonpert}.  The relatively coarse lattice
spacing was chosen in order to 
allow for larger physical volumes,                              
and thus reduce the corrections to scaling
at any given lattice size.

\begin{table}
\centerline{
\begin{tabular}{ll}
\hline
lattice & measurements \\
\hline
$16^3$  &  200000 \\
$24^3$  &  200000 \\
$32^3$  &  300000 \\
$40^3$  &  350000 \\
$48^3$  &  400000 \\
$64^3$  &  400000 \\
\hline
\end{tabular}
}
\caption[a]{
The set of simulations for $\beta_G=5$. All simulations have
been performed at $x=0.112706$, $\beta_H=0.362835$, and later reweighted
to the estimated position of the infinite volume critical point:
 $x=0.11331593$, $\beta_H=0.36288657$.
}
\label{tab:stat}
\end{table}

\subsection{Locating the critical point \la{critpoint}}

Let us first recall how one locates the critical point in the
case of a one-dimensional, rather than a two-dimensional, parameter space,
such as in the case of the spin models in \eq\nr{O(N)}. 
Here the critical point
is known to occur at $h=0$, and the only parameter which remains to be
found is $\beta_c$.

In this case the procedure commonly used is the ``intersection of
Binder cumulants'' \cite{binder1}. It is based on the following general
idea. Consider the system in a finite box of given geometry (say,
cubic) with given boundary conditions (say, periodic). Consider any
observable (for example, magnetization), averaged over the system. Make
a Boltzmann ensemble of configurations, for each configuration measure
this observable and thus construct its probability distribution. Then,
if the system is exactly at the critical point and scaling is
valid, the {\em form} of this probability distribution should be
independent of the system size; only its scale will be changing. Thus
any characteristics such as 
those in \eqs(\ref{M^4}), (\ref{E^3}), designed to be
sensitive to the form of probability distribution but not to the
rescaling of the observable, will behave in the following typical way:
when plotted as a function of a parameter (for spin systems, as a
function of $\beta$) for several lattice sizes, all plots intersect at
the value of the parameter that corresponds to the critical point.
This provides a convenient way to locate it.

This approach can easily be generalized to the case of a
two-dimensional parameter space. The main
idea remains the same: the critical point is a point where the form of
two-dimensional distributions, such as those in Fig.~\ref{352t-all}, does not
depend on the system size, up to a possibly nonorthogonal linear
transformation. To locate the critical point, one now needs {\em two}
characteristics. For example, if one has somehow found the $M$-like
direction, one can consider 
$\langle (\Delta \tM)^4 \rangle /
\langle (\Delta \tM)^2 \rangle^2$ and $\langle (\Delta
\tM)^3 \rangle / \langle (\Delta \tM)^2\rangle^{3/2}$. 
The former is sensitive to a deviation from the critical point along the
(continuation of) the first order transition line, while the latter is
sensitive to a deviation across the line. Finding the intersection of 
these cumulants for
two lattice sizes now implies solving a system of two equations
for two variables.

The procedure just described is completely general (however, it remains
to be understood how to find the $M$-like direction; see Sec.~\ref{MandE}) 
and does not
depend on any conjecture about the universality class of the critical
point. However, it appeared not to be very practical, the main
stumbling block being its sensitivity to corrections to scaling, which
are in our case non-negligible, as demonstrated by Fig.~\ref{352t-all}.

A modification of this approach which is more stable against deviations
from scaling, relies on a
conjecture about the universality class of the critical point. Indeed,
if we expect the critical point to belong to the 3d Ising universality
class, we know that in the scaling limit 
$\langle (\Delta \tM)^4
\rangle / \langle (\Delta \tM)^2 \rangle^2 = 1.604(1)$
\cite{Bloete}, $\langle (\Delta \tM)^3 \rangle / \langle (\Delta
\tM)^2\rangle^{3/2} = 0$.
By solving these equations, one can find 
the apparent location of the critical point for each lattice
size separately. The consistency of the conjecture about the
universality class can be checked later: the lattice size dependence of
the position of the apparent critical point should follow the known
correction to scaling behaviour.

As a variant, one can use the whole probability distribution $P(\tM)$, 
which is known quite precisely for the 3d Ising model,
rather than its moments, and require that the apparent critical point
for the given lattice size be a point where $P(\tM)$ matches most
favourably the Ising $P(M)$, say, by the $\chi^2$ criterion. This
method has been used in \cite{wilding} for locating the liquid-gas
critical point.

The method we chose to use in practice is as follows. 
Assume that we have determined an $M$-like direction, as explained
in the next Section. 
To compute the position of the apparent critical point for a given
lattice size we have reweighted the data for the corresponding
lattice (Table~\ref{tab:stat}) to a trial value of $(x,\beta_H)$
which translates to $(\beta_H,\beta_R)$ according to \eq\nr{betar}, 
computed the probability distribution of the
$M$-like observable $P(\tM)$ and tuned $(x,\beta_H)$
so that

1. the two peaks of $P(\tM)$ are of equal weight,

2. the ratio of the peak value of $P(\tM)$ (the average
height of the two peaks) to $P(\tM)$ at the minimum between
the peaks equals the corresponding ratio for the 3d Ising
model at the critical point \cite{TsyBlo}:
\be
P_\rmi{max}/P_\rmi{min} = 2.173(4).
\ee
The criterion based on the ratio $P_\rmi{max}/P_\rmi{min}$ appeared
to be less sensitive to asymmetric corrections, which are most
pronounced at the tails of $P(\tM)$, than the usual one
based on the fourth order cumulant in \eq\nr{M^4}.

\begin{figure}[t]

\centerline{
\epsfxsize=8cm\epsfbox{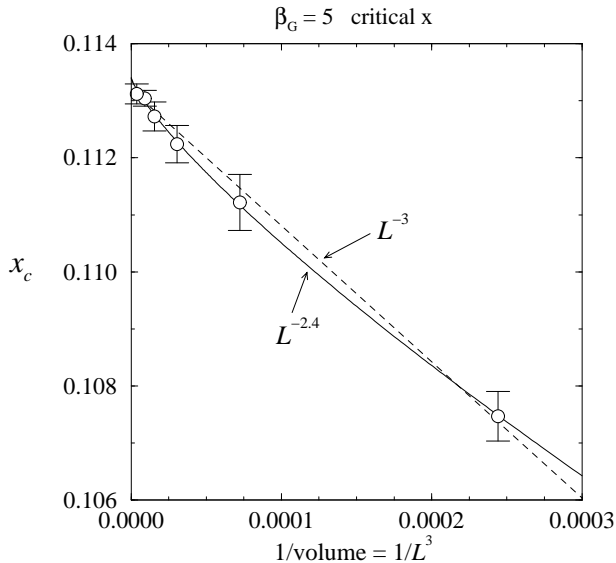}
}

\vspace*{-0cm}

\caption[a]{
The reweighted $x_c$ from simulations
at $\beta_G=5,x = 0.112706$, as a function of the volume.
}
\la{xcrit_betag5}
\end{figure}

The resulting dependence of the apparent position of the critical
point on the lattice size is shown in Fig.~\ref{xcrit_betag5}.
It follows nicely the law based on Ising-type corrections to
scaling, the deviation of $x_c$ from the limiting value behaving as
\be
x_c(L) - x_c(\infty) \propto L^{-(\Delta+1)/\nu} \approx L^{-2.4},
\ee
where $\Delta = 0.52(4)$ is the universal correction to scaling
exponent for the 3d Ising universality class.  Thus the determination
of the critical point based on the expected Ising-like properties is
completely consistent.  However, the statistical errors of the
datapoints are large enough so that a regular $L^{-3}$ -behaviour
cannot be ruled out, either.  Nevertheless, the variation of the
infinite volume critical point is quite small, and it has a negligible
effect on the analysis below.  The infinite volume result, determined
by the $L^{-2.4}$-fit, is $x_c(\infty) = 0.1133(25)$.  In the
following analysis we always reweight the data to the critical point
$x_c=0.11331593,\,\beta_H=0.36288657$.  Due to the strong correlations
in coupling constants both have to be fixed to a high numerical
precision.

\subsection{Determining M-like and E-like observables \la{MandE}}

We observed that even the problem of locating the critical point, 
to say nothing of further quantitative analysis, depends on
finding the $M$-like direction in the space of observables. In
Fig.~\ref{352-1000}, this is done by letting the $M$-like direction go
along the probability distribution, and 
taking the $E$-like to be orthogonal to it.
The result appears to be very encouraging, by the eye, when compared with
the 3d Ising distribution in Fig.~\ref{p51}(c). This approach relies on
the fact that the distribution is extremely elongated, and becomes even
more so with growing lattice size, fluctuations of $\tM$ growing
much faster than those of 
$\tE$: $\langle (\Delta \tM)^2
\rangle / \langle (\Delta \tE)^2 \rangle \propto L^{(\gamma -
\alpha)/\nu} \approx L^{1.8}$.

The determination of the $M$-like and $E$-like directions can now be put on
a quantitative basis as follows. Take the probability distribution
$P( S_\rmi{hopping}, S_{(\phi^2-1)^2} )$. Compute the fluctuation
matrix and find its eigenvectors. The larger eigenvalue will give
$\langle (\Delta \tM)^2 \rangle$, the smaller one 
$\langle (\Delta \tE)^2 \rangle$, while the corresponding eigenvectors
will give the $M$-like and $E$-like directions.

The procedure just described is identical to the one used in
\cite{alonso} and completely ignores the possibility that the $M$-like
and $E$-like directions can be nonorthogonal 
in the original basis chosen \cite{MerminRehr,wilding}.
However we have found that both this simplistic procedure and its
generalization to a larger space of observables, which is discussed
below, work extremely well for our system, while the more sophisticated
approach of \cite{wilding} runs into serious difficulties. This seems
to be an interesting point that deserves some discussion, as it
probably means that asymmetric corrections to scaling play a more
prominent role in our system than in the liquid-gas models (see also
Sec.~\ref{sec:asymm}). 

The method
of \cite{wilding} employs the matching of the probability distribution of a
linear combination of the two basic observables to $P(M)$ known from
the 3d Ising model at criticality, to find the $M$-like direction
simultaneously with the apparent critical point (performing a search in
3-dimensional space: two parameters for a trial critical point, plus
one parameter for a trial $M$-like direction). After that, the $E$-like
direction is found by matching the distribution of a trial linear
combination of observables to the 3d Ising $P(E)$.

One of the key statements of \cite{wilding} is that the pronounced
asymmetry of two-peak probability distributions of various observables
at the critical point can be attributed to the fact that they are
actually mixtures of $\tM$ and $\tE$, while the
distribution of $\tM$ itself comes out completely symmetric,
within the accuracy of the simulation. One can, however, raise the
following question: is it not possible that the perfect symmetry of
$P(\tM)$ emerges as an artefact of the procedure (optimization of
its matching to the exactly symmetric $P(M)_\rmi{Ising}$)?

\begin{figure}[p]

\centerline{
(a)
\epsfysize=6.2cm \epsfbox[36 40 539 468]{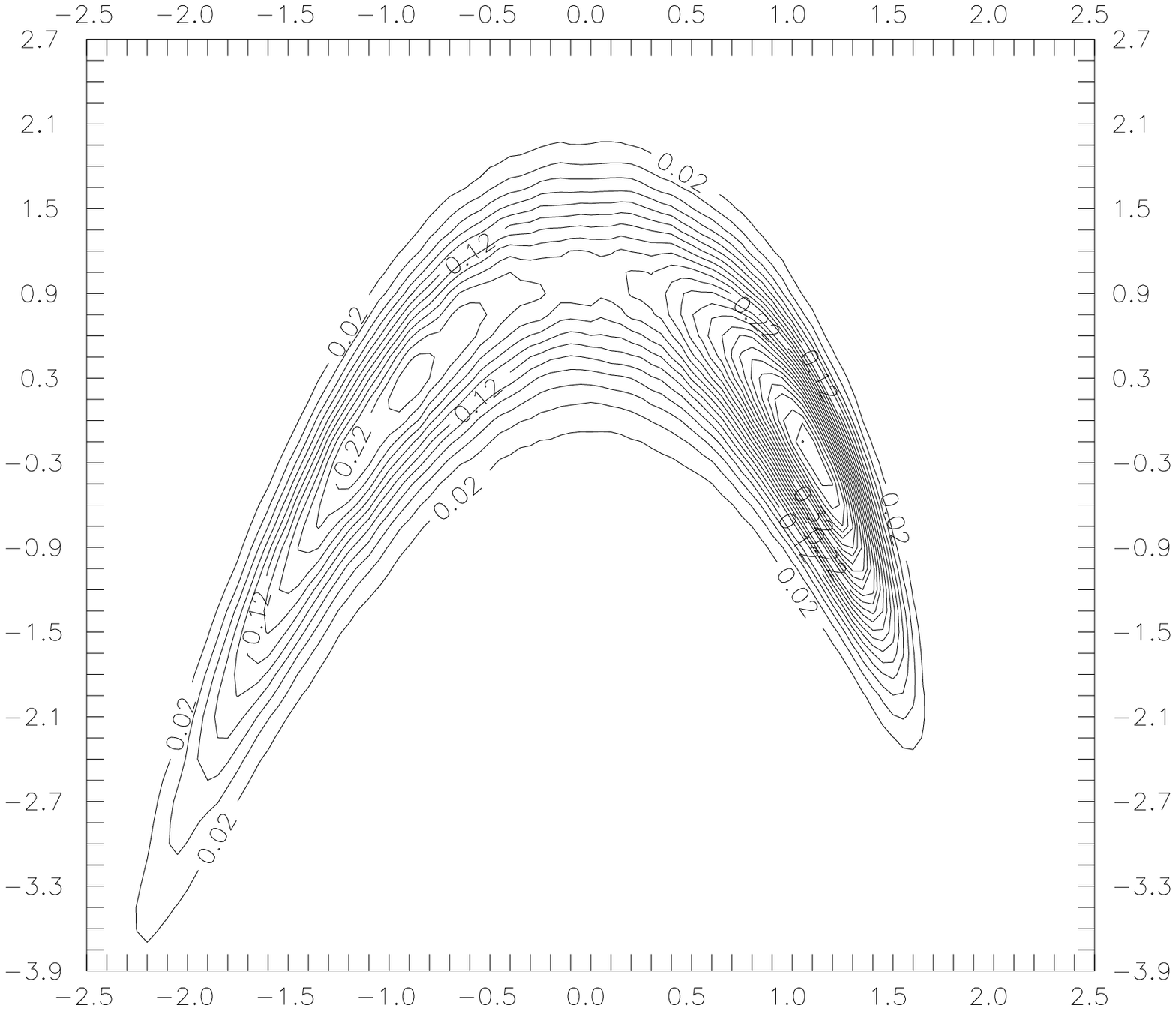}
\epsfysize=5.2cm \epsfbox{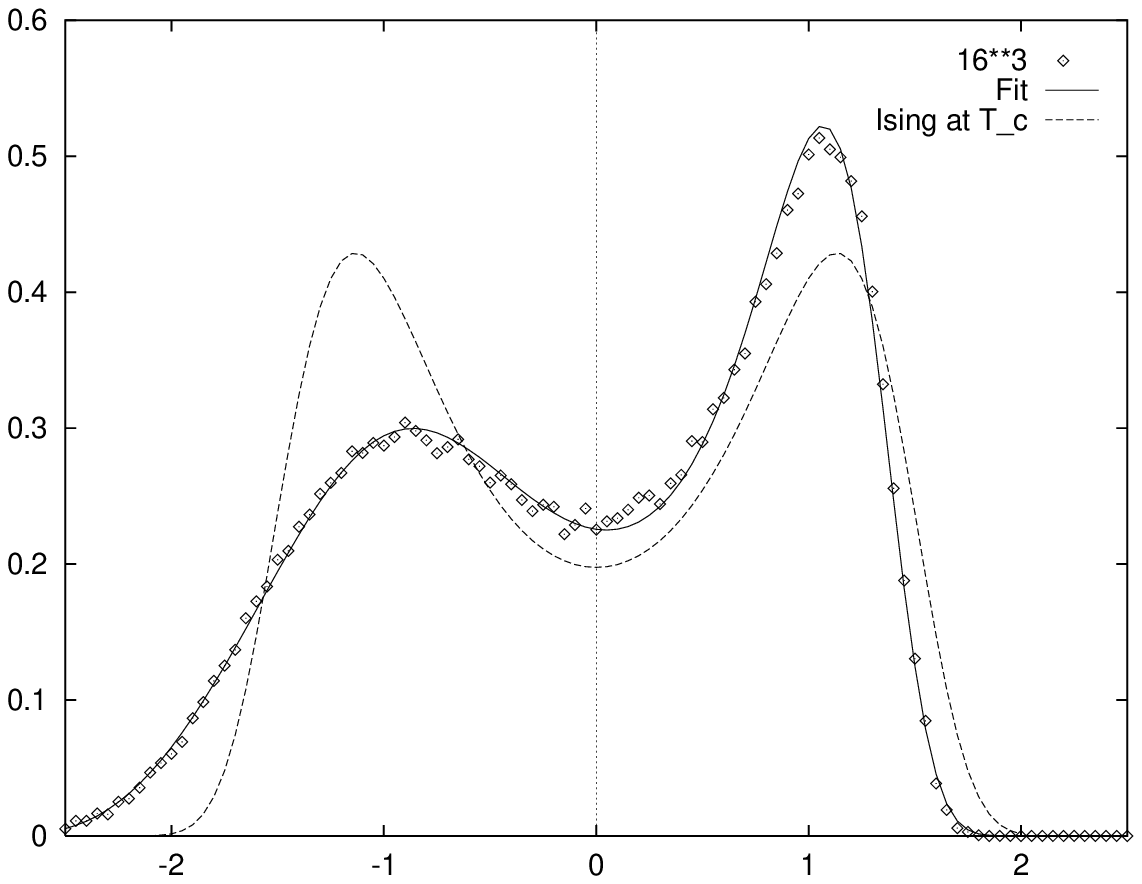}
}
\centerline{
(b)
\epsfysize=6.2cm \epsfbox[36 40 539 468]{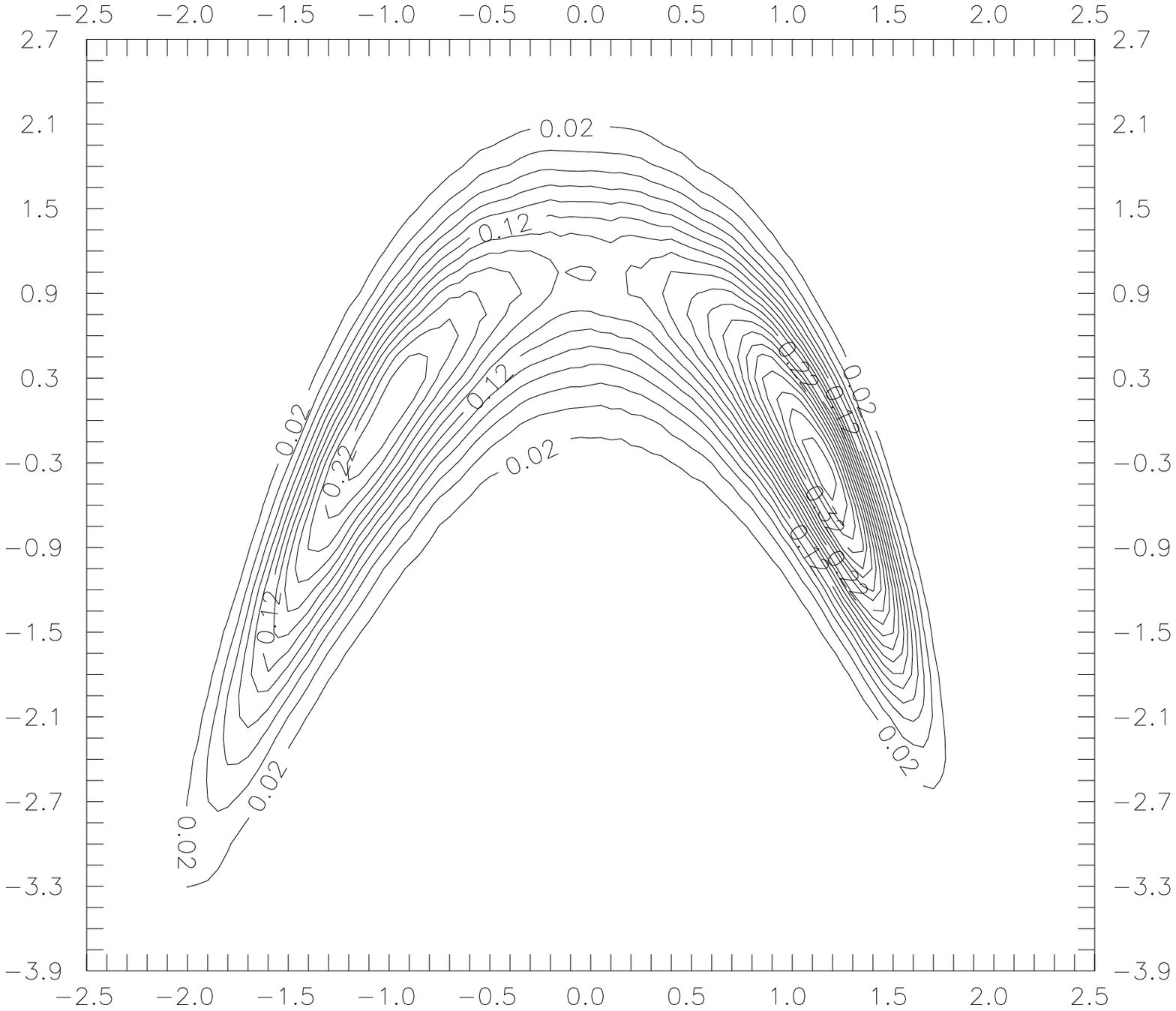}
\epsfysize=5.2cm \epsfbox{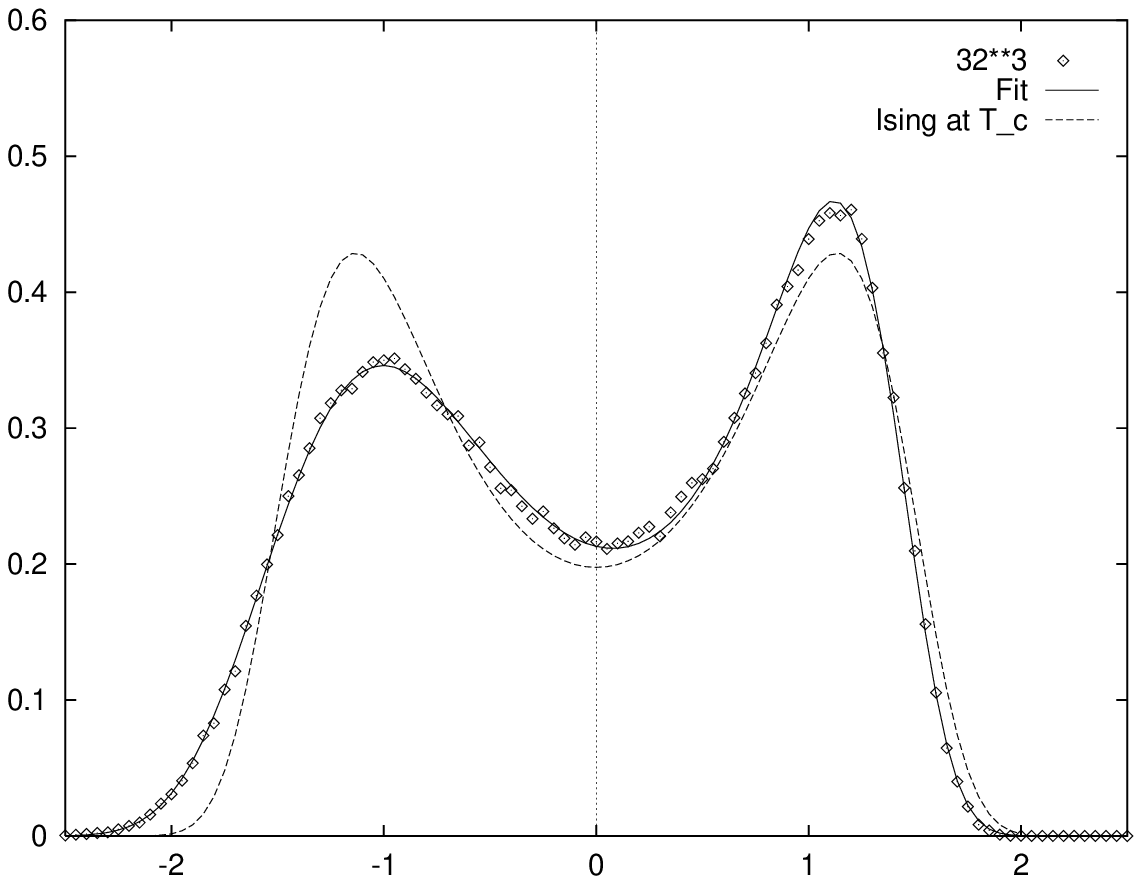}
}
\centerline{
(c)
\epsfysize=6.2cm \epsfbox[36 40 539 468]{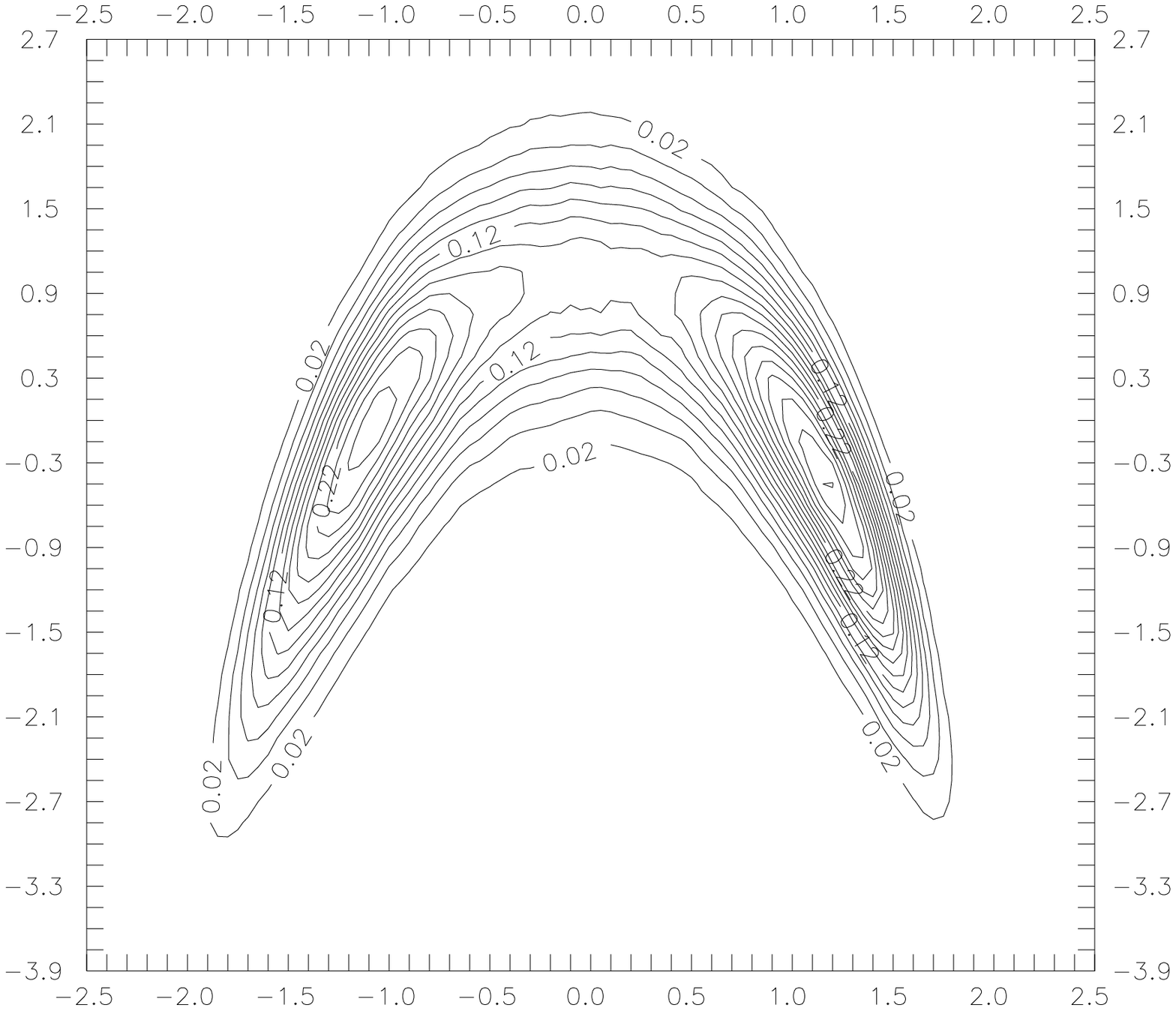}
\epsfysize=5.2cm \epsfbox{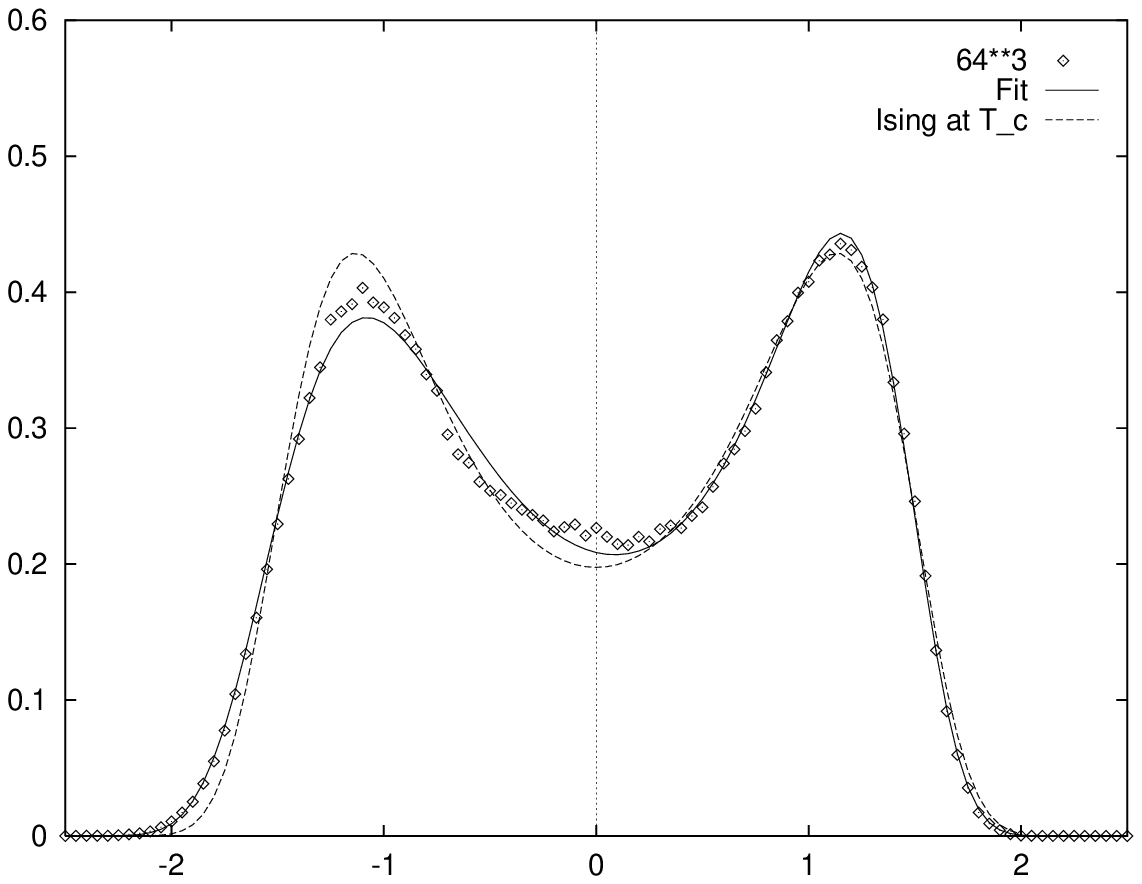}
}
\caption[a]{
The probability distributions $P(\tM,\tE)$
({\em left\/}) and $P(\tM)$ ({\em right\/}) at the infinite volume
critical point, for the volumes 
(a) $16^3$, (b) $32^3$, (c) $64^3$. It is seen
how the distribution becomes more symmetric for
increasing volumes.  
The $\tM$ and $\tE$ directions have been found
with a 6-dimensional fluctuation matrix analysis, 
see Sec.~\ref{sec:extending}.
}
\la{volumes}
\end{figure}

This problem can be also put as follows. We have the two-dimensional
probability distribution for our system, as in Fig.~\ref{352t-all}(b),
now as a function of four parameters: the trial critical point and two
trial directions for $\tM$ and $\tE$ (not necessarily  
orthogonal). On the other hand, we have the corresponding 
$P(M,E)_\rmi{Ising}$ for
the Ising model at criticality, Fig.~\ref{p51}(c), with its projections
$P(M)_\rmi{Ising}$ and $P(E)_\rmi{Ising}$. 
The question is, is it a good idea to look for the
$M$-like and $E$-like directions by requiring that just the
one-dimensional projections of Figs.~\ref{352t-all}(b) and \ref{p51}(c)
onto the horizontal and vertical axes match each other? 
Should not one rather match the whole distributions?

Obviously, in the absence of deviations from scaling it would make
little difference whether to match the whole distributions or their
one-dimensional projections: both methods would converge to the same
result, corresponding to a perfect matching. The problem, however,
becomes nontrivial when there are deviations from scaling, especially
the asymmetric ones. Concretely, we have found that for our system, 
an application of the procedure in \cite{wilding}
appeared to be completely misleading.

\begin{figure}[t]

\centerline{
\epsfysize=6.2cm \epsfbox[36 40 539 468]{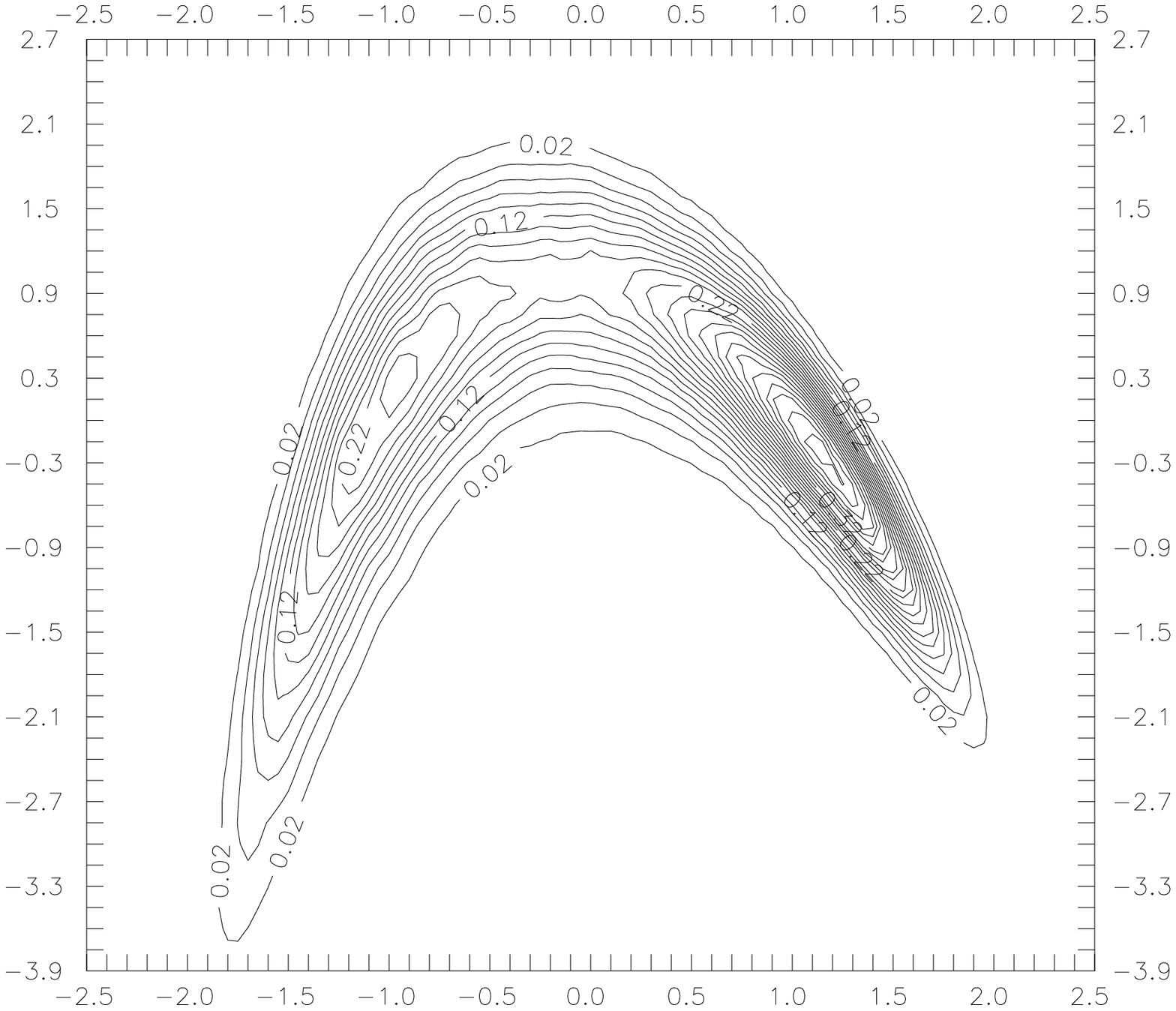}
\epsfysize=5.2cm \epsfbox{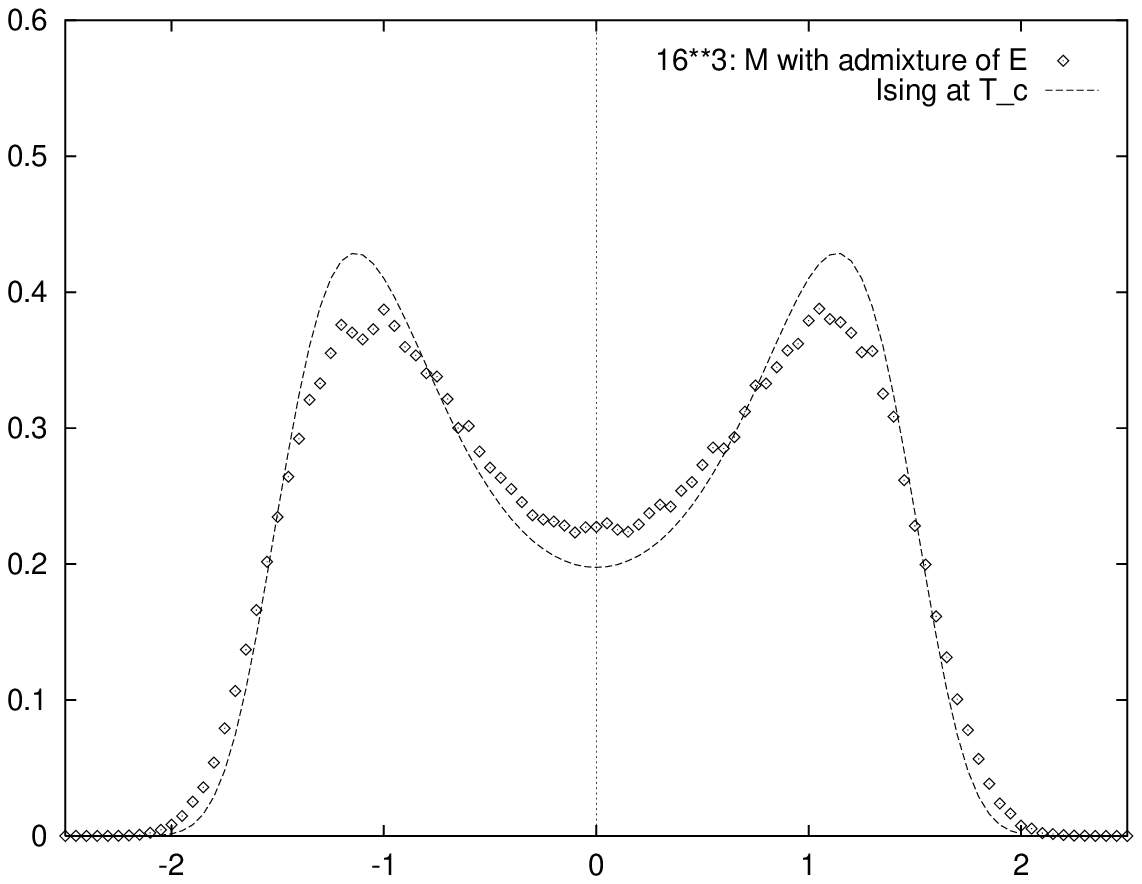}
}
\caption[a]{
This figure shows that it is possible to find a direction for
 $\tM$ that provides a perfectly symmetric $P(\tM)$,
but only at the price of reducing the symmetry of
 $P(\tM,\tE)$. The data are the same as in
Fig.~\ref{volumes}(a).
}
\la{wrong2d}
\end{figure}

The observation is that 
for practically achievable lattices, there is a significant asymmetry in
two-dimensional distributions, as seen in Fig.~\ref{352t-all}(b). 
This asymmetry appears to be unremovable
(more precisely, only a relatively small part of it can be
removed) by any choice of (nonorthogonal) directions for $\tM$ and $\tE$.  
This can be understood when one notices that one of
the characteristic features of this asymmetry is the difference of
areas under the two peaks of the probability distribution (see also
Figs. \ref{volumes}), which cannot be cured by
any linear transformation, even nonorthogonal, as such a
transformation keeps the ratio of areas invariant.

Thus if we try to symmetrize the two-dimensional distribution
(or match it to the Ising form Fig.~\ref{p51}(c)), a considerable
asymmetry remains and, notably, $P(\tM)$ comes out considerably
asymmetric (Fig.~\ref{volumes}, right). At the same time, matching
$P(\tM)$ to $P(M)_\rmi{Ising}$ easily finds the ``$M$-like''
direction that ensures a perfect matching and thus symmetric
$P(\tM)$ (Fig.~\ref{wrong2d}). 
But this optimization of the symmetry of the  
one-dimensional projection is achieved at the price of greatly {\em
reducing}, rather than improving, the symmetry of the two-dimensional
histogram as a whole and thus should be
considered completely misleading!

Thus we have found the following important differences between our
system and the liquid-vapour models \cite{wilding}:

1. Our system demonstrates non-negligible asymmetric corrections to
scaling. These show up in two-dimensional distributions and cannot be
removed by any choice of $\tM$ and $\tE$. As a consequence,
asymmetries of various one-dimensional distributions are mostly
caused by them, and not so much by the admixture of $\tE$, as in
\cite{wilding}.

2. In our system, we do not find much evidence of the possible
nonorthogonality of the $M$-like and $E$-like directions. Deviations
from orthogonality, if any, can be safely neglected, in clear
distinction from \cite{wilding} where they played quite a prominent
role.

\begin{figure}[t]

\centerline{
\epsfysize=6.2cm \epsfbox[36 40 539 468]{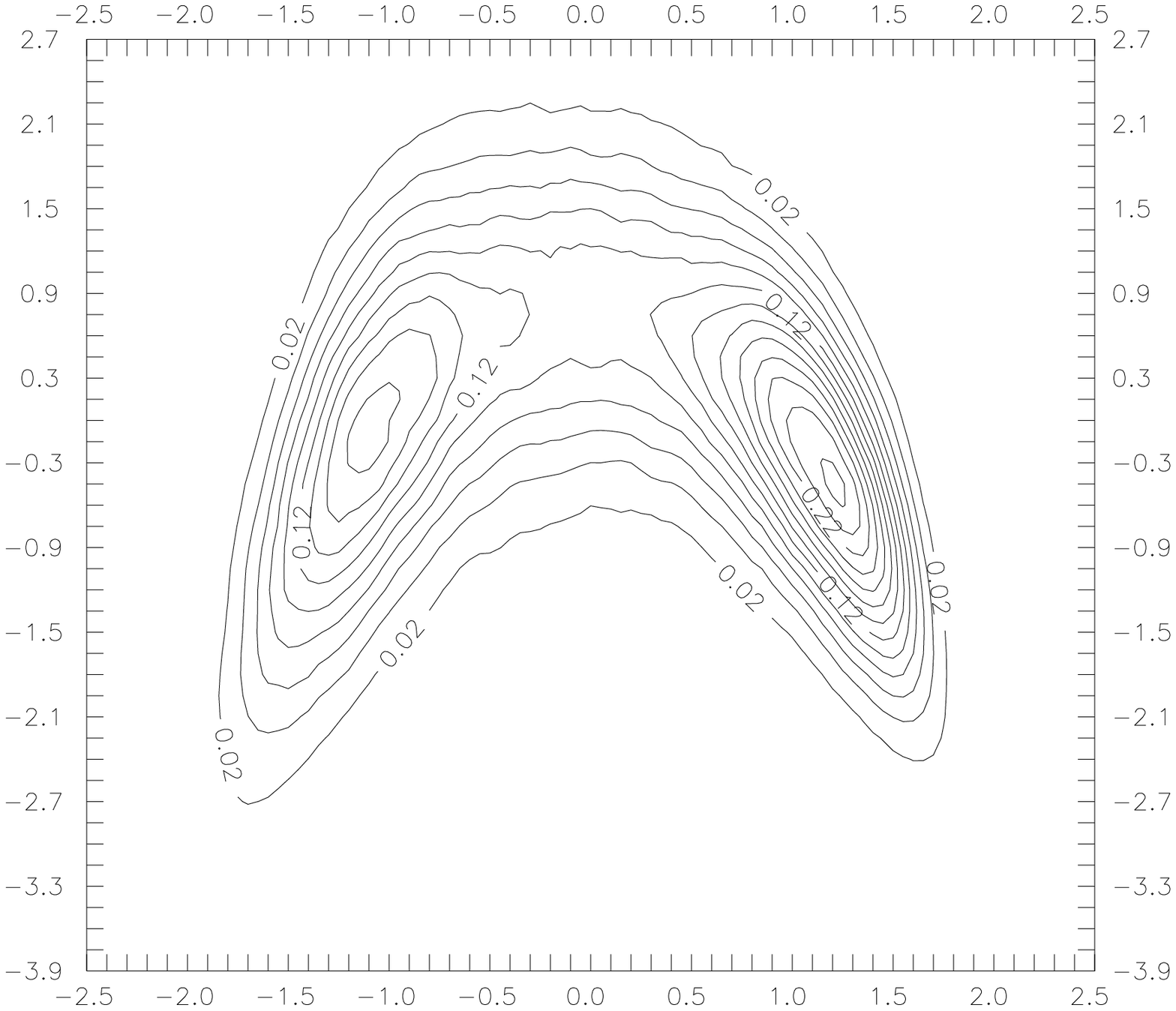}
\hspace*{5mm}
\epsfysize=6.2cm \epsfbox[36 40 539 468]{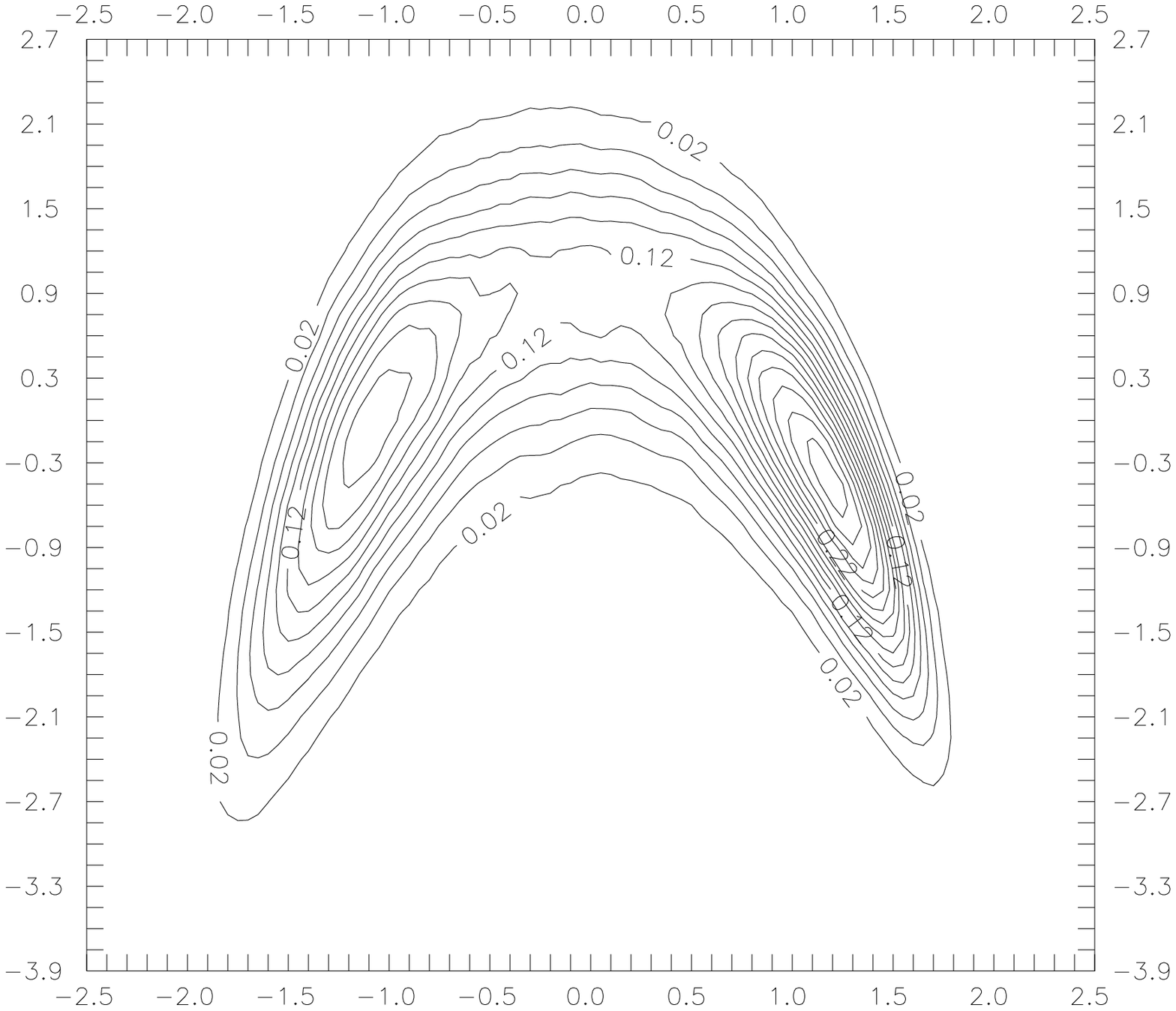}
}

\caption[a]{
The dependence of the probability distribution $P(\tM,\tE)$
on the number of observables,
for a $\beta_G=5$, $64^3$ lattice.
{\em Left:}
The diagonalized fluctuation matrix for two observables
 $S_\rmi{hopping}$, $S_{(\phi^2-1)^2}$,
see \eq\nr{latticeaction}.
{\em Right:}
The same for four observables,
 $S_G$, $S_\rmi{hopping}$, $S_{\phi^2}$, $S_{(\phi^2-1)^2}$.
It is seen that the distribution becomes sharper, or thinner,
as the basis is expanded. Using six observables leads to
a still sharper distribution, see Fig.~\ref{volumes}(c).}
\la{basis}
\end{figure}

In conclusion, 
after having tried four methods for
determining the $M$-like and $E$-like
directions, 

(a) finding the eigenvectors of the fluctuation matrix \cite{alonso},

(b) matching $P(\tM)$ to $P(M)_\rmi{Ising}$,
 $P(\tE)$ to $P(E)_\rmi{Ising}$ \cite{wilding},

(c) matching $P(\tM,\tE)$ to $P(M,E)_\rmi{Ising}$,

(d) maximizing the symmetry of $P(\tM,\tE)$,

\noindent
we arrived at the conclusion that the method (a) works best for our
system, method (b) appears to be misleading, and methods (c) and (d)
produce results consistent with (a), while being much more difficult to
implement and use. One of the tricky points is the multidimensional
minimization of the difference of two Monte Carlo generated
two-dimensional probability distributions, which is typically a very
noisy function (the problem is alleviated by first smoothing the
histograms, then minimizing their difference). Another stumbling block
of the method (c) is that a seemingly harmless manifestation of
deviations from scaling --- the excessive thickness of  
$P(\tM,\tE)$ compared with $P(M,E)_\rmi{Ising}$ --- has a
very strong effect on their difference, in terms of $\chi^2$, making it
impossible to achieve a good matching.

\subsubsection{Extending the space of observables}\la{sec:extending}

The main observation so far was that while the form
of the probability distribution at the critical point comes out
strikingly similar to that of the 3d Ising model, as demonstrated by
Figs.~\ref{352t-all}(b,c), 
there are still
differences (asymmetry and 
thickness) that are decreasing with growing
lattice size, but relatively slowly, so that, for example, the
elimination of 
the thickness would require prohibitively large lattice
sizes.

The situation can be considerably improved by
further generalizing the procedure of determining the $M$-like and
$E$-like observables. The reasoning behind this is as follows. If we
consider any arbitrary observable (say, $S_\rmi{hopping}$), it behaves
at the critical point more or less like magnetization, and its
probability distribution also looks very similar to the distribution of
magnetization, the main feature being the double-peak structure.
However, it shows a certain asymmetry, which eventually goes down to
zero with growing lattice size. This can be understood as a consequence
of the fact that we expect any observable to behave at the critical
point as a sum of $M$-like, $E$-like and regular contributions. Their
dependence on the lattice size is different and is governed,
correspondingly, by $L^{\gamma/\nu}$, $L^{\alpha/\nu}$ and $L^0$. On
large lattices the magnetic contribution is always dominating, so any
given observable starts behaving as the $M$-like.

{}From this point of view, when we study the joint distribution of two
observables and find the $M$-like and $E$-like directions as the
primary axes of the corresponding very elongated fluctuation ellipse,
we are actually finding the $E$-like direction as a linear combination
of observables in which the dominating $M$-like terms cancel each
other. Thus the consideration of two-dimensional distributions provides
a way to disentangle the dominant ($M$-like) and subdominant terms.
However, it becomes clear that within this approach the $E$-like
observable will collect all subdominant terms, both actually $E$-like
and regular.

Thus one arrives at the idea that a further separation of $E$-like and
regular contributions could be achieved by generalizing the procedure
to more observables than two. The hope is to ``purify'' the $\tE$
(for $\tM$ there is little difference, as it outweighs everything
else by orders of magnitude anyway).

\begin{figure}[t]

\centerline{
\epsfysize=6.2cm \epsfbox[36 40 539 468]{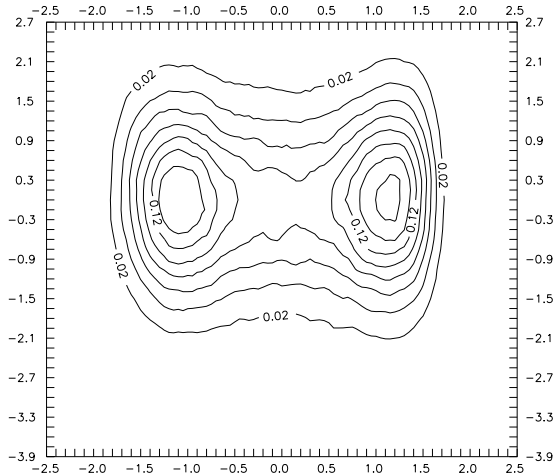}
}
\caption[a]{
The joint probability distribution of observables corresponding
to the largest (horizontal axis) and second largest (vertical
axis) eigenvalues of the $4\times 4$ fluctuation matrix
of the terms in \eq\nr{latticeaction}.}
\la{2dPlotWithReg}
\end{figure}

To begin with, we have considered the 4-dimensional space of
observables, these observables being the four terms in the action
in \eq\nr{latticeaction}. Diagonalizing the $4 \times 4$ matrix
$\langle(S_i-\langle S_i \rangle) (S_j-\langle S_j \rangle)\rangle$ for
the $64^3$ lattice $(\beta_G=5)$ resulted in the eigenvalues
and -vectors shown in Table~\ref{tab:eigenvalues}.
We observe a pronounced hierarchy of
eigenvalues, similar to the previously considered case of two
observables, Fig.~\ref{352-1000}(a).
The largest eigenvalue corresponds, as expected, to
$\tM$. However, $\tE$ turns out to correspond, somewhat
surprisingly, to the {\em smallest} eigenvalue, rather than to the
second largest one, while the two eigenvalues in the middle correspond
to regular directions. This is substantiated both by analysing the
dependence of the eigenvalues on the lattice size and by looking at the
joint probability distributions of various pairs of the 4 observables
corresponding to the 4 eigenvectors. The joint distribution of projections
onto eigenvectors corresponding to the largest and to the smallest
eigenvalue is depicted in Fig.~\ref{basis}(right); projections onto
eigenvectors corresponding to the largest and to the second largest
eigenvalue produce a strikingly different pattern,
Fig.~\ref{2dPlotWithReg}, signalling that the second largest eigenvalue
does indeed correspond to a regular observable: its fluctuations are
Gaussian-like and independent from those of $\tM$.

It is evident in Fig.~\ref{basis}
that the extension of the space of observables from two-
to four-dimensional does indeed considerably reduce the deviation of
$P(\tM,\tE)$ from the 3d Ising scaling form. The most
notable effect is the reduction of the excessive 
thickness of
$P(\tM,\tE)$. Now the question is, what happens if we
further extend the space of observables, having in mind that if one
wants to sort out $\tE$ as well as possible, one would like it to
correspond to an eigenvalue which is not the smallest one (as the
smallest one is just collecting all unresolved contributions). Thus we
have added two additional observables: the sum of the absolute values
of the Higgs field, and the analog of the hopping term, where the Higgs
matrices have been replaced by SU(2) matrices, dividing out the
length of the Higgs field:
\be
\tilde S_R \equiv \sum_\bfx R(\bfx), \quad
\tilde S_L \equiv \sum_{\bfx,i}
\fr12\tr V^\dagger(\bfx)U_i(\bfx)V(\bfx+i) , \la{otherops}
\ee
where $\Phi(\bfx)=R(\bfx)V(\bfx)$, $R\ge0,\,V\in$ SU(2).

Now the energy eigenvalue appears to be the fourth of six, in
descending order, and we observe further significant reduction of
difference between $P(\tM,\tE)$ and $P(M,E)_\rmi{Ising}$,
as seen in Fig.~\ref{volumes}(c).

One could continue extending the space of observables (in principle,
there are infinitely many gauge invariant operators to be considered),
but these six operators seem to be enough for our purposes.

The coefficients of the different eigenvectors
in terms of the original operators in 
\eqs\nr{latticeaction},\nr{otherops}, 
together with the 
eigenvalues, are shown in Table~\ref{tab:eigenvalues}.
We observe the following:

1. While the eigenvectors corresponding to the
three largest eigenvalues are relatively stable with respect to
an increase in the number of basis vectors, the $E$ direction
changes considerably. However, 
the final critical distributions, critical indices, etc, 
are quite stable.  

2. The largest eigenvalue, the magnetic one, is about 4 orders of 
magnitude larger than the next largest, for the volumes used.     

3. The second largest eigenvalue consists almost solely of the
plaquette term of the action, and conversely, the plaquette
term contributes significantly only to the second eigenvalue.
Thus, the plaquette term is practically decoupled from the other modes.

4. At very large volumes the energy eigenvalue will overtake the two
regular eigenvalues above it and become the second largest one
(Sec.~\ref{sec:critindex}).  However, for the range of volumes studied 
here, the hierarchy shown in Table~\ref{tab:eigenvalues} was 
preserved.

\begin{table}[t]
\centering
\begin{tabular}{llllllll}
\hline
direction & $\lambda$ & $c_1$ & $c_2$ & $c_3$ & $c_4$ & $c_5$ & $c_6$ \\ \hline
\multicolumn{8}{c}{4 operators} \\ \hline
$M$ & 1.28$\times10^{10}$ & 0.05142 & 0.72590 & -0.68564 & -0.01808 
& -- & -- \\
regular & 8.51$\times10^{5}$ & 
0.9965 & 0.008 & 0.083 & 0.0049 & -- & -- \\
regular & 2.59$\times10^{5}$  &
-0.066 & 0.6877 & 0.7227 & 0.0185 & -- & -- \\
$E$ & 1.75$\times10^3$ & 
-0.0027 & 0.0004 & -0.0262 & 0.99965 & -- & -- \\
\hline
\multicolumn{8}{c}{6 operators} \\ \hline
$M$ & 1.33$\times10^{10}$ &
0.0505 & 0.7133 & -0.67375 & -0.01777 & -0.1646 & -0.0853 \\
regular & 8.52$\times10^5$ & 
0.9954 & 0.010 & 0.087 & 0.0055 & 0.0082 & -0.037 \\
regular & 2.81$\times10^5$ & 
-0.078 & 0.655 & 0.6876 & 0.0262 & 0.136 & -0.271 \\
$E$ & 1.32$\times10^5$ & 
0.024 & 0.233 & 0.033 & -0.1052 &  0.450 & 0.855 \\
regular & 4.05$\times10^3$ & 
1$\times10^{-5}$ & 
-0.0914 & -0.241 & -0.217 & 0.836 & -0.433 \\
regular & 73 & 
-2$\times10^{-5}$ &  9$\times10^{-5}$
& -0.0816 & 0.9700 & 0.229 & 0.0019 \\ \hline
\end{tabular}
\caption[a]{\protect
The eigenvalues $\lambda$ and the coefficients $c_i$ for
the diagonalized directions, in terms of the operators in 
\eqs\nr{latticeaction},\nr{otherops}.
Here the volume is $64^3$, $\beta_G=5$,
and the data have been reweighted to the infinite volume critical point.
}
\la{tab:eigenvalues}
\end{table}

\begin{figure}[t]

\centerline{
\epsfxsize=7.2cm\epsfbox{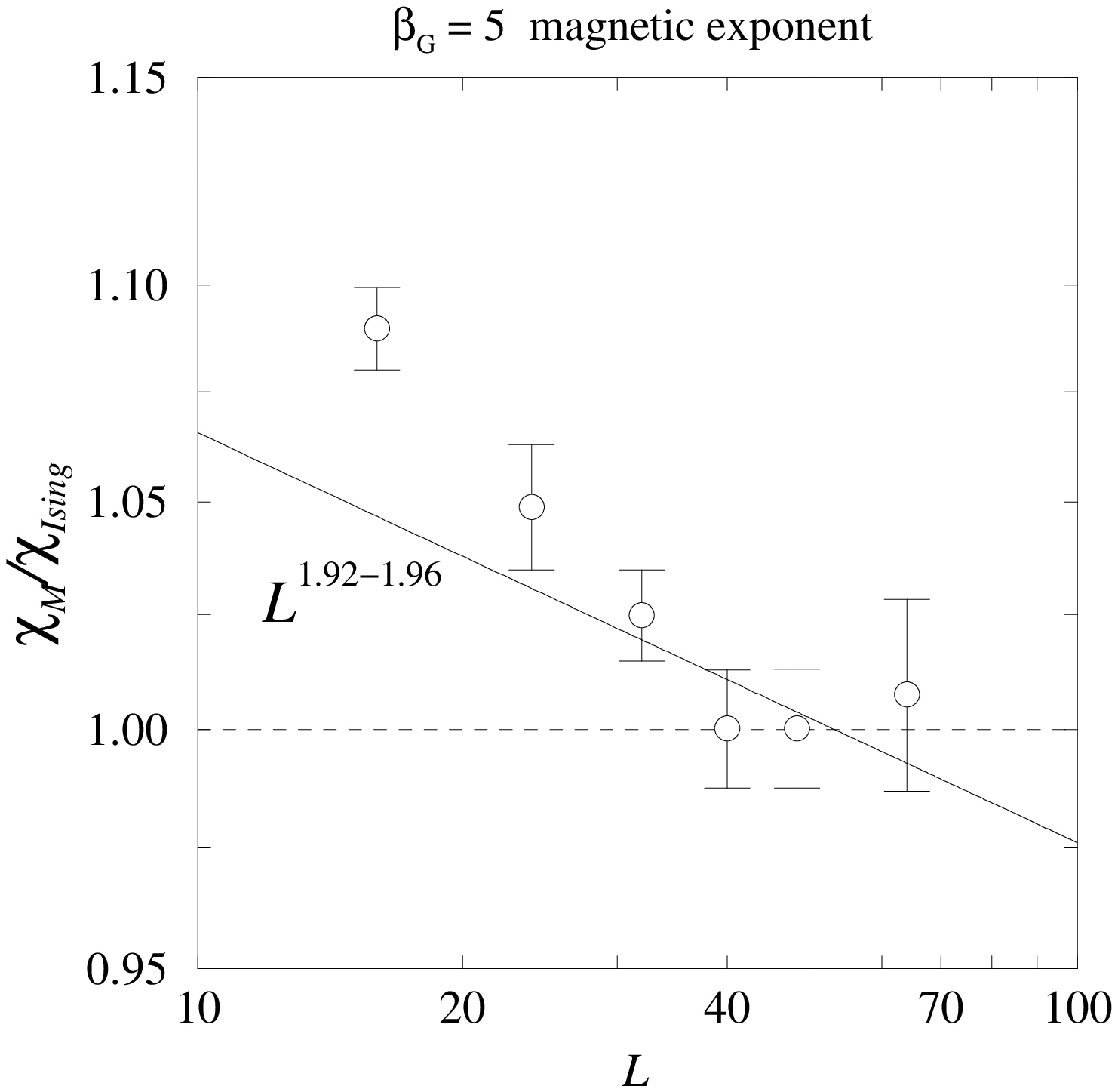}\hspace{3mm}
\epsfxsize=7.1cm\epsfbox{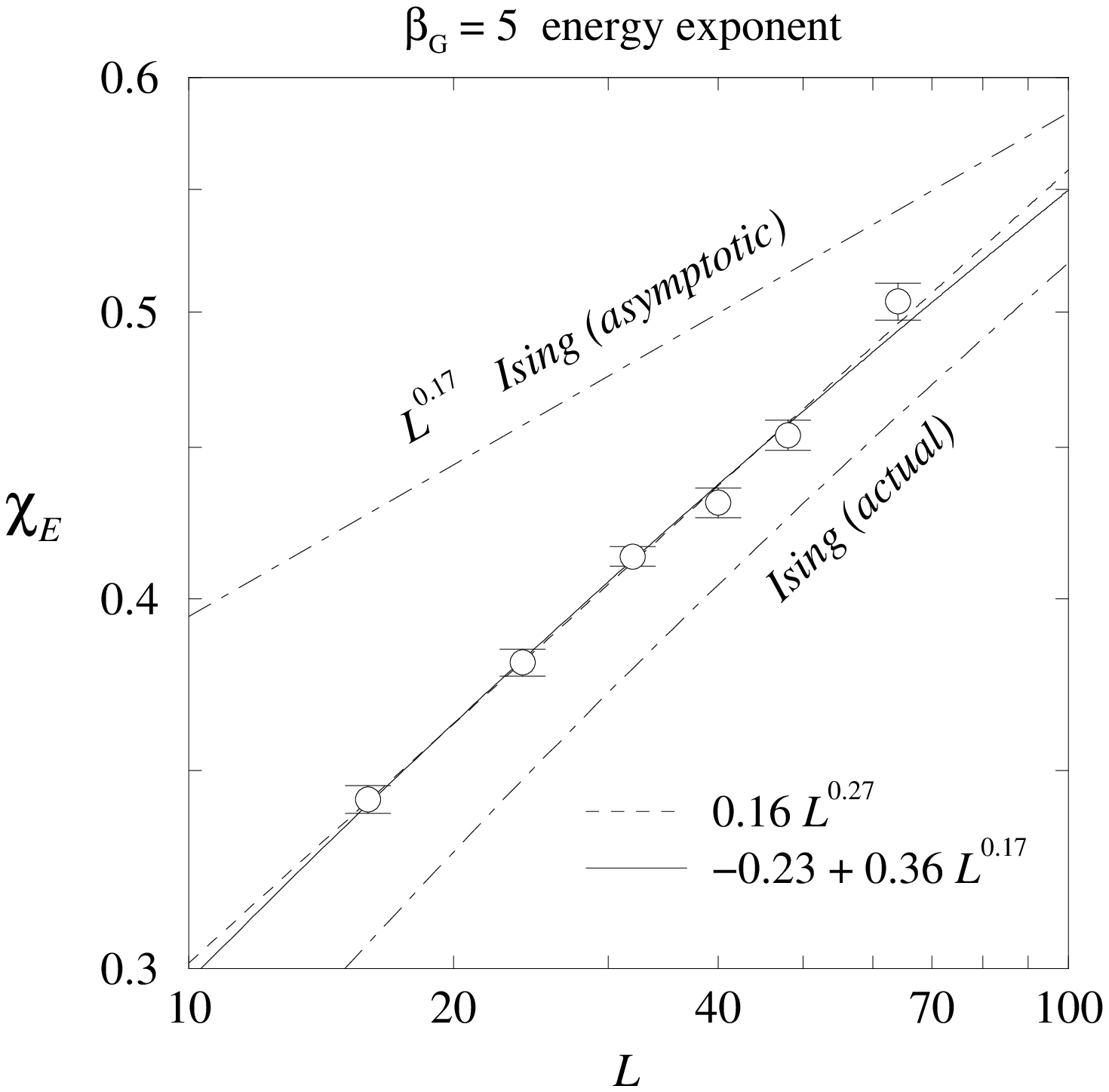}}

\caption[a]{{\em Left:}
the magnetic susceptibility $\chi_M$
divided by the Ising scaling law $\mbox{const}\times L^{1.96}$.
The corresponding critical exponent is
$\gamma/\nu=1.92(3)$, whereas Ising model has
$\gamma/\nu = 2 - \eta \approx 1.96$.
It is seen that at larger volumes the results for $\chi_M$ are
consistent with the Ising model.
The absolute value of $\chi_M$ is $\sim 5\times 10^4$ at $L=64$.
{\em Right:}
the energy susceptibility $\chi_E$.
Note that the absolute value is much smaller
than for $\chi_M$.
Two different fits to the datapoints are shown. 
It is seen that the behaviour is
consistent with that of the Ising model
(which is described by $\chi_E= -11.1 + 14.6 L^{0.17}$
\cite{HaPinn} and is shown here up to an arbitrary
overall factor, so only its slope is relevant).
O(N) models with $N\ge 2$ have a negative 
exponent for $\chi_E$ and are thus excluded.}
\la{fig:chis}
\end{figure}

\begin{figure}[thb]

\centerline{
\epsfxsize=7.8cm\epsfbox{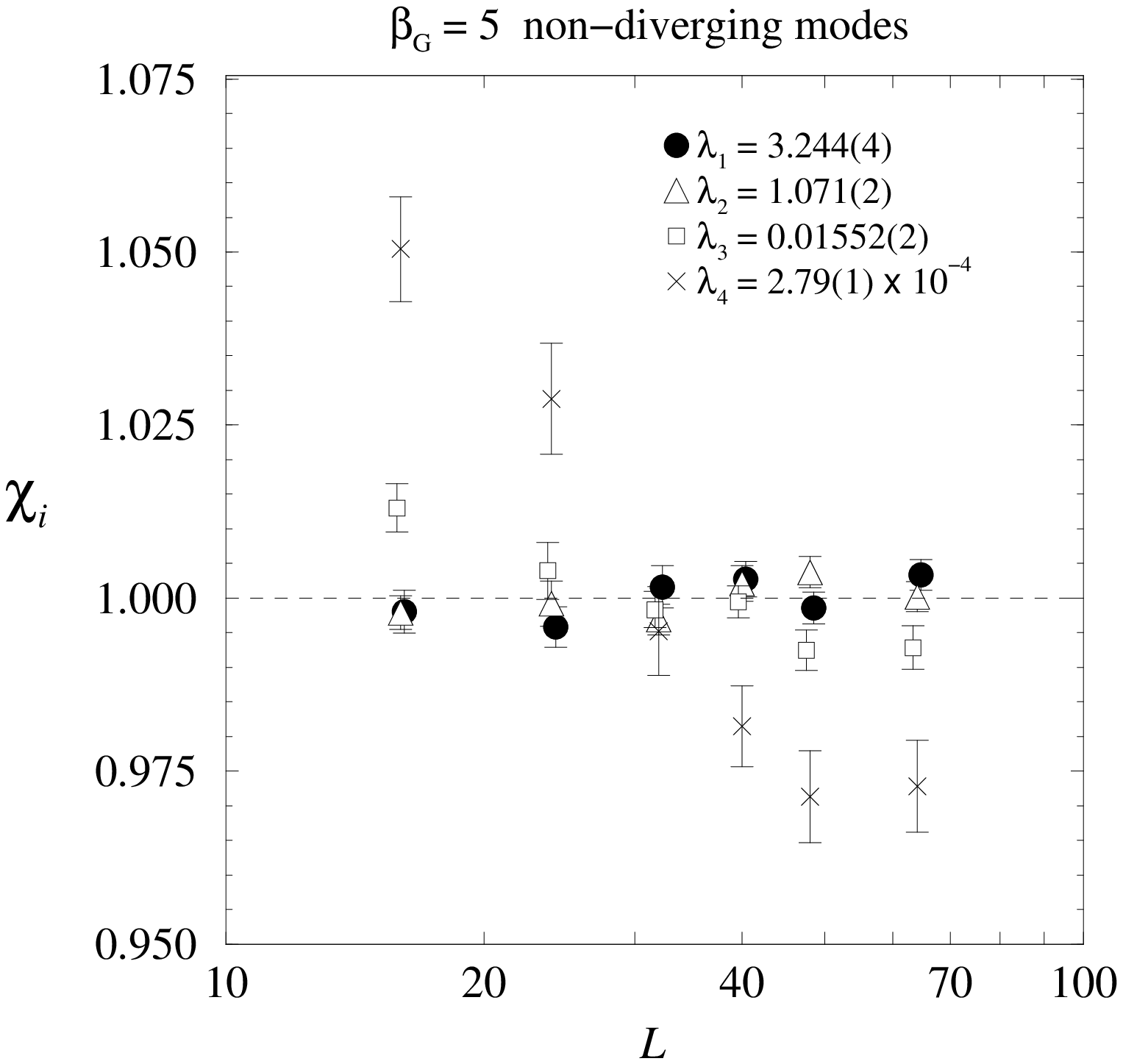}}

\caption[a]{The regular eigenvalues, divided by the volume,
as a function of the
lattice size. The normalization is arbitrarily
chosen such that the average of the values shown is 1.0;
the absolute values are as indicated by the numbers.
It is seen that the eigenvalues are
constant with a very good accuracy (note the scale
of the $y$-axis). The smallest eigenvalue has the largest
volume dependence as it is contaminated by higher states.}
\la{fig:regular}
\end{figure}

\subsection{Critical indices}
\la{sec:critindex}

Now that the $M$- and $E$-like directions have been determined,
one can find the critical indices,
using the finite size scaling formulas in \eq\nr{gamma}. 
The scaling has to be studied at the infinite volume 
critical point $x_c(\infty)$, 
whose determination was discussed in Sec.~\ref{critpoint}.
There is a small dependence of eigenvectors on the lattice size, due to
corrections to scaling; we take a fixed set of eigenvectors
(corresponding to the largest lattice, $64^3$) and compute the second
moments of the corresponding projections, for a set of lattice sizes. The
dispersion of $\tM$ grows approximately as $L^{4.92}$, the dispersion
of $\tE$ grows as $L^{3.27}$, and those of the remaining projections
grow as $L^3$, as shown in Figs.~\ref{fig:chis},
\ref{fig:regular}. (An additional volume factor enters due to
observables being sums over the lattice, without dividing by the volume).
The apparent value of $\alpha/\nu \approx 0.27$ deviates notably from
the Ising asymptotic value 0.17, but just the same effect is observed
for the Ising model itself, for similar lattice sizes.  This is explained
by the presence of a negative regular background in $\chi_E$
\cite{HaPinn}, as shown in Fig.~\ref{fig:chis}.

\begin{figure}[t]

\centerline{
\epsfxsize=7.8cm\epsfbox{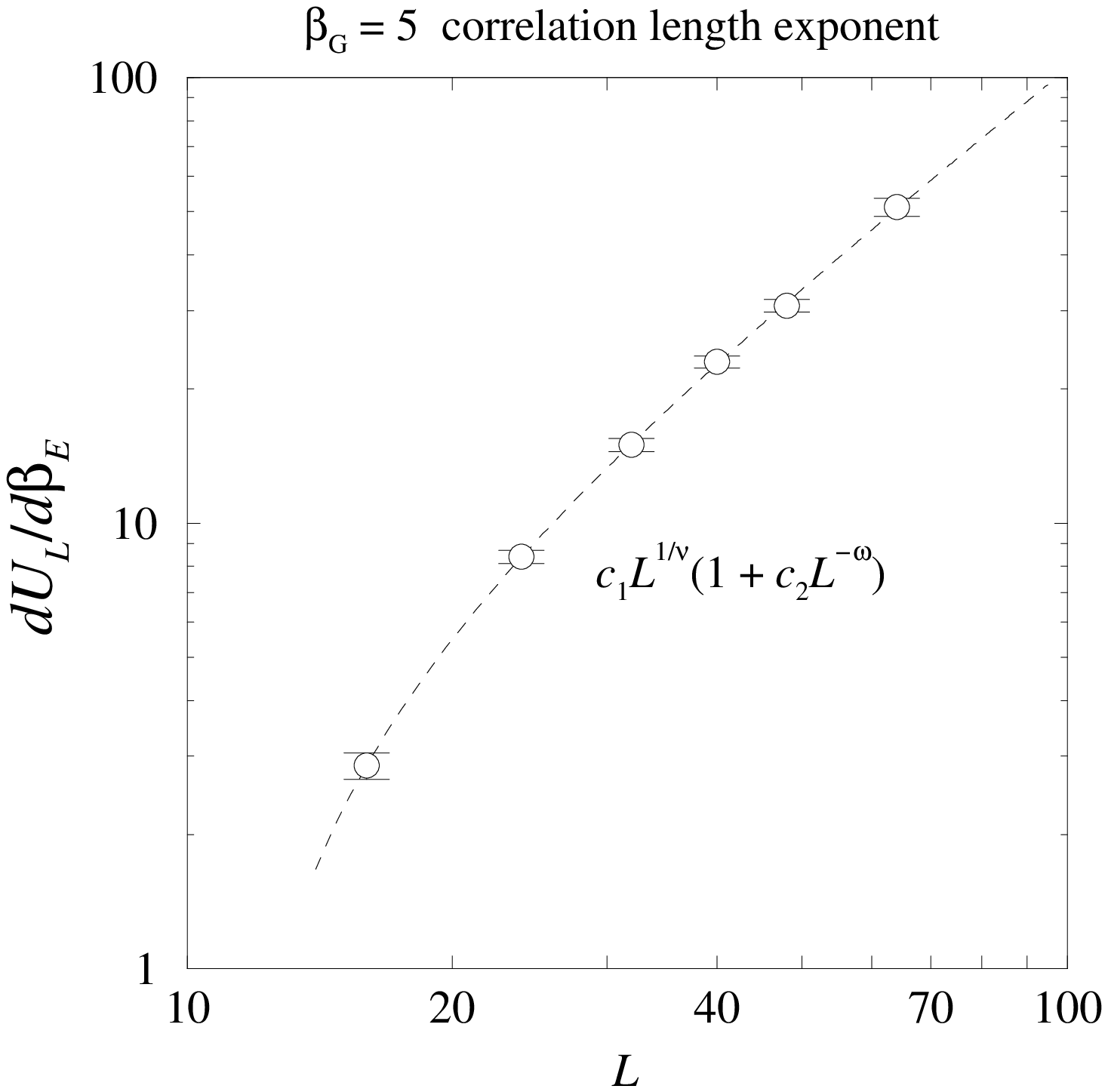}}

\caption[a]{ The slope of the Binder cumulant $U_L$ at the critical
point in the $E$-like direction versus the lattice size.  The dashed
line is a 4-parameter fit to the data.} \la{fig:nu}
\end{figure}

\vspace{4mm}
In addition to the critical exponents $\gamma/\nu$ and $\alpha/\nu$,
which are related to the second moments of the distributions $P(M)$ and
$P(E)$, we have also determined 
the correlation length critical exponent $\nu$.  In models with
an exact symmetry $M\leftrightarrow -M$ (like the Ising model),
finite observables like the Binder cumulant 
\be
 U_L \equiv 1 - \fr13 \fr{\langle M^4\rangle}{\langle M^2\rangle^2}
\la{bc}
\ee
can be assumed, near the critical point, to be regular functions 
of $\xi/L$, the ratio of the correlation length
to the system size.  As $\xi \sim t^{-\nu}$, the exponent $\nu$
can be obtained from
the slope of $U_L$ at the critical point \cite{binder1}:
\be
  \fr{\partial U_L}{\partial t} \propto L^{1/\nu}\,.
\la{dUL}
\ee
The SU(2)+Higgs theory lacks the explicit $M\leftrightarrow -M$ 
-symmetry, and we use the $M$- and $E$-like eigenvectors in the 
analysis.  Furthermore, we substitute $M \rightarrow \Delta M = M -
\langle M\rangle$ in the Binder cumulant.  Let us now denote by $\beta_E$ 
the coupling constant of the $E$-like eigenvector: that is, we
formally extend the lattice action in \eq\nr{latticeaction}  
to the form $S + 
\beta_E E$, where $\beta_E = 0$ at the critical point.  
We then obtain from \eqs\nr{bc} and \nr{dUL} that 
\be
   \fr{\partial U_L} {\partial \beta_E} =
   (1-U_L) \left[ 
	\langle E \rangle + 
	\fr{\langle (\Delta M)^4 E\rangle}{\langle (\Delta M)^4\rangle} -
	2\fr{\langle (\Delta M)^2 E\rangle}{\langle (\Delta M)^2\rangle}
	\right]\,.
\ee
This expression is readily evaluated using the $M$- and $E$-like
eigenvectors at the critical point.  The results are shown in
Fig.~\ref{fig:nu}.  With these volumes the corrections to scaling are
still substantial, and the points do not fall on a straight line on a
log-log plot.

Taking into account the corrections to scaling, we fit the 
data with the 4-parameter ansatz
\be
  \fr{\partial U_L} {\partial \beta_E} =
  c_1\,L^{1/\nu} \,\left( 1 - c_2\,L^{-\omega} \right)\,.
\ee
However, with only 6 points and relatively large statistical errors
(when compared with, say, the Ising model simulations \cite{hpv}) 
the error range in the correlation length critical exponent becomes
rather large: the result is $\nu = 0.63(17)$.  If we lock the 
correction to scaling 
exponent $\omega$ to the central value $\omega = 1.7$
and perform a three-parameter fit, the result becomes $\nu = 0.63(3)$.

The value of $\nu$ is completely compatible 
with the Ising model (but also with O(2)). 
However, in the Ising model the correction to scaling 
exponent is $\omega = 0.8$,
which does not fit the data well.  This is very likely due to the asymmetry
in the $P(M,E)$-distributions of the smallest volumes (Fig.~\ref{volumes}).
In order to observe the Ising-type corrections to scaling one should
use volumes which are large enough so that the asymmetric corrections
to scaling have become subdominant.  This is discussed in more
detail in the next section.

\begin{figure}[t]

\centerline{ 
\epsfxsize=8cm\epsfbox{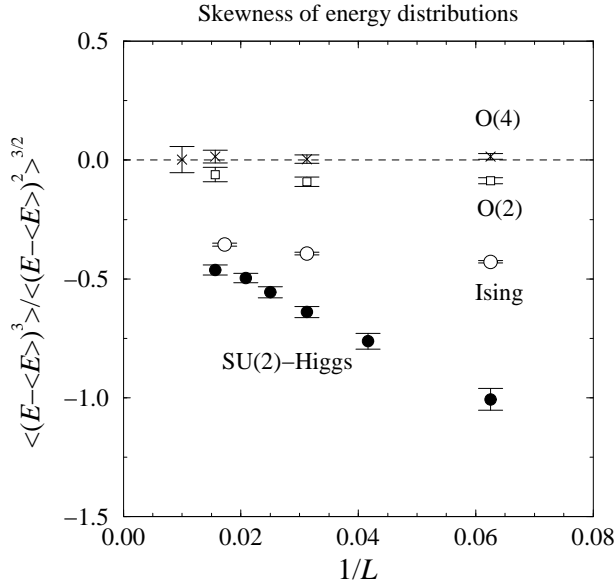}}

\caption[a]{
The skewness of $P(\tE)$ for 3d SU(2)+Higgs and different
spin models, versus the inverse lattice size.
}
\la{fig:eskew}
\end{figure}


\vspace{4mm}
Apart from the second moments $\chi_E$, $\chi_M$, we have also measured
the third moment of $\tE$ and the corresponding skewness ratio in \eq\nr{E^3}.
The results are shown in Fig.~\ref{fig:eskew}. They are consistent with
those of the Ising model, but differ from O(2) and especially from
O(4), in which case the skewness of $P(\tE)$ is very small.

\section{Dependence of asymmetry on the volume}
\la{sec:asymm}

Significant asymmetry effects were observed in the previous
analysis of the SU(2)+ Higgs theory. Here we study them in 
some more detail. We consider 
the distributions $P(\tM)$ at the critical point, as
shown in Fig.~\ref{volumes}(right), and attempt to obtain a
quantitative description of how they approach the Ising shape with
growing lattice size.

It is well-known \cite{Wegner1} that the leading corrections to scaling
in an exactly symmetric system, such as the Ising model itself, show
the universal behaviour governed by $L^{-\omega}$, $\omega \equiv
\Delta/\nu \approx 0.8$ (see also Sec.~\ref{critpoint}). 
However, the {\em asymmetric} corrections,
which are also present in our system, have their own critical exponent
$\omega_5$, which is different from $\omega$ and has attracted much
less attention in the literature (to our knowledge, it has never been
studied before in the framework of Monte Carlo simulations). 
Including the operators $\phi^5,\phi^6$ in \eq\nr{isingaction},
the exponent $\omega_5$ has been computed within the
$\varepsilon$-expansion up to order $\varepsilon^3$
\cite{Wegner2}--\cite{ZhangZia}, and within the renormalization group
framework \cite{NewRie}. Quoting from \cite{ZhangZia},
\begin{equation}
\omega_5 = 1 + {11 \over 6} \varepsilon -
            {685 \over 324} \varepsilon^2 +
            {107855 + 103680 \zeta(3) \over 34992} \varepsilon^3
            + O(\varepsilon^4).
\end{equation}
This series behaves poorly at $\varepsilon=1$, resulting in $1 + 1.83 -
2.11 + 6.64 \ldots$. An attempt to improve the situation using Pad\'e
approximants produces the sequence 2.83, 1.85, 2.32, in orders
$O(\varepsilon)$, $O(\varepsilon^2)$, $O(\varepsilon^3)$ respectively,
leading to the estimate that $\omega_5 \gsim 1.5$ \cite{ZhangZia}. This
estimate has been confirmed by the computation within the
renormalization group \cite{NewRie}, which resulted in $\omega_5 =
2.4(5)$. Thus it is generally believed (see, e.g., \cite{Anisimov}),
that $\omega_5 \approx 2.1$ (the average of the last two Pad\'e
values). This implies that the asymmetric corrections, going as
$L^{-\omega_5}$, should die out very rapidly when a critical point is
approached, not only faster than the leading symmetric corrections
$(L^{-\omega})$, but also faster than the subleading ones
$(L^{-2\omega})$, and thus be of no practical importance anywhere 
near the critical point. This is probably the reason why $\omega_5$ is
rarely discussed in the literature.

However, we {\em do} observe significant asymmetric contributions,
and it is interesting to see what kind of implications there are
for $\omega_5$.
For this purpose we need to quantify the deviation of $P(\tM)$
from the Ising scaling form (Fig.~\ref{volumes}, right). As has been
found in \cite{TsyBlo}, the scaling form of $P(M)$ for the 3d Ising
model in a cubic box with periodic boundary conditions is described
extremely well by the following simple approximation:
\be
  P(M) \propto \exp \biggl\{ - \biggl({M^2 \over M_0^2}-1\biggr)^2
  \biggl(a{M^2 \over M_0^2}+c\biggr) \biggr\},
 \label{IsiShape}
\ee
where $M_0$ is the position of the magnetization peak, and the
universal constants $a$ and $c$ are
\ba
 a & = & 0.158(2),   \label{AandC} \\
 c & = & 0.776(2).   \nonumber
\ea
The generalization of \eq\nr{IsiShape} for our case must also include
odd powers of $M$, and can be written down as follows:
\be
 P(\tM) \propto e^{-V_\rmi{eff}}, \quad
  V_\rmi{eff} = (x^2-1)^2 (ax^2+bx+c)-hx, \quad x= \frac{\tM-\tM_1}{\tM_0}.
 \label{AsymShape}
\ee
This $V_\rmi{eff}$ can be considered as a special parametrization
of a general double-well polynomial of up to sixth order in $\tM$.
The parameters 
$\tM_0$ and $\tM_1$ are responsible for rescaling and shifting
the distribution as a whole. Two of the remaining four parameters,
 $b$ and $h$, characterize the asymmetry of $P(\tM)$.

One of them, $h$, is sensitive to what may be called the {\em superficial
asymmetry}: an asymmetry caused by a small deviation of parameters of our
system from the true critical point across the first order line,
which is, in the case of the Ising model, equivalent to an application of
a small external field. This parameter is also sensitive to the asymmetry
caused by statistical errors in the relative heights of the two peaks,
which emerge on larger lattices as a consequence of the growth of
the tunnelling time with the lattice size.

On the other hand, the parameter $b$ characterizes what may be called
the {\em genuine asymmetry}, and is responsible for the difference in the
peak widths that remains after we make them equally high by slightly
shifting the system across the phase transition line (that is, in
the $h$-like direction).

\begin{figure}[t]

\centerline{
\epsfxsize=8cm\epsfbox{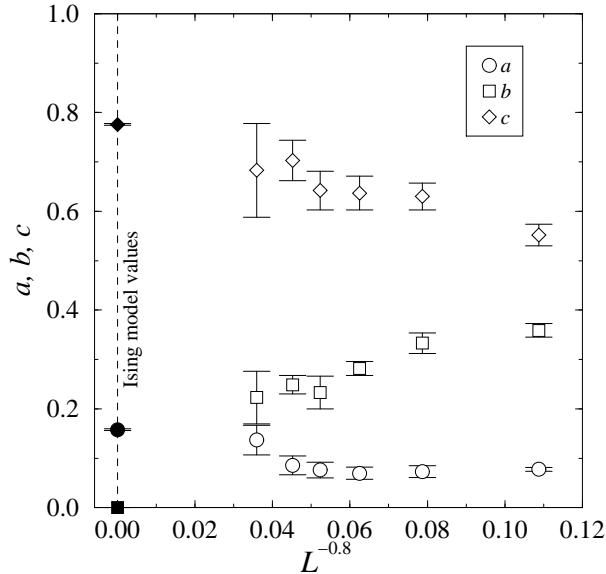}
}
\caption[a]{\protect
The parameters $a$, $b$, $c$ that determine the shape of
$P(\tM)$ according to \eq\nr{AsymShape}, 
as a function of the lattice size ($L=16\ldots64$).
The values in \eq\nr{AandC} for
the Ising model in the scaling limit are marked on the left edge
of the plot.
}
\la{fig:abc}
\end{figure}

Thus we will be interested in three parameters: $a$, $b$ and $c$.
Making a 6-parameter fit to our data, we observe
(Fig.~\ref{volumes}, right) that the ansatz in \eq\nr{AsymShape}
is indeed able to provide a sufficiently good approximation.
The quality of the approximation turns out not to be
exactly as excellent as in the case of the 3d Ising model
(one observes, for example, that the fit tends to go a little
bit above the top of the higher peak), but quite sufficient for
our purposes.

The results for $a$, $b$ and $c$ are shown in Fig.~\ref{fig:abc}.
One observes that all three parameters go in the directions
of the corresponding Ising limits, with growing lattice size.

The parameter $c$ falls reasonably well on a straight line which
corresponds to the standard correction to the scaling exponent
 $\omega \approx 0.8$. It is interesting to note that it
approaches the scaling limit from below, while in the
simple cubic Ising model it approaches the same limit from above
\cite{TsyBlo}. 

The parameter $a$ behaves less nicely but is also consistent with
 $\omega \approx 0.8$ for larger lattice sizes.
As for the parameter $b$, which is expected to behave as $L^{-\omega_5}$,
it does indeed decrease with growing lattice size, but much
more slowly than implied by the generally accepted high
value of $\omega_5$. In fact, $b$ seems to go down {\em more slowly}
than $L^{-0.8}$, rather than going as $L^{-2.1}$
(the data in Fig.~\ref{fig:abc} are best fitted
by $L^{-0.4} \ldots L^{-0.5}$)!

The origin of this contradiction remains unclear. It might be that our
lattices are still too small, and we have not yet reached the
asymptotic regime where asymmetric corrections behave as
$L^{-\omega_5}$. On the other hand, something might be missing in the
theoretical treatment of asymmetric corrections to scaling. This
question certainly deserves further study.

\section{Summary of the results}\la{sec:summary}

The aim of this paper was to study the universality properties
of the endpoint of the line of first order phase transitions in
the 3d SU(2)+Higgs gauge theory. Qualitatively, the result was obvious
when comparing the two-dimensional near-endpoint
distribution of this theory
in E- and M-like variables, shown in Fig.~\ref{352t-all}(b), 
with the corresponding distribution for the 3d Ising model shown in
Fig.~\ref{352t-all}(c): the distributions look extremely
similar. The bulk of this paper was devoted to putting this
similarity on a strict quantitative basis.

Indeed, 
the two main discrepancies between Figs.~\ref{352t-all}(b) and
\ref{352t-all}(c) -- the asymmetry and the 
thickness of
Fig.~\ref{352t-all}(b) --  can be removed by going to the infinite
volume limit and by using a larger basis of observables.
The effect of the volume variations is shown in Fig.~\ref{volumes}.
It is seen that as the volume is increased, the distribution looks
more and more like that of the Ising model, Fig.~\ref{spinmodels}(a).
The effect of the choice of basis is shown in Fig.~\ref{basis}.
It is seen that the 
thickness is removed as one goes from two
to four observables, and even more as one goes to six
observables (Fig.~\ref{volumes}(c)). The fact the Fig.~\ref{volumes}(c)
agrees with Fig.~\ref{spinmodels}(a) and not with
Figs.~\ref{spinmodels}(b),(c), is our main result.

Furthermore, the basis of six observables allows to construct a
$6\times6$ correlation matrix, which is then diagonalized.
It turns out that two of the eigenvalues show critical
behaviour, whereas four are regular. The largest
eigenvalue corresponds to $\chi_M$. The susceptibilities
$\chi_M,\chi_E$ are shown in Fig.~\ref{fig:chis} as a function
of the volume. The behaviour is clearly consistent with
that of the Ising model. For the energy exponent $\chi_E$
even the fact that scaling violations are large at moderate
volumes is reproduced. An O(4) model with a negative
exponent for $\chi_E$ is excluded. 

The remaining four
regular exponents are shown in Fig.~\ref{fig:regular}.
They show no critical behaviour as a function of the volume.

Apart from the second moments $\chi_E,\chi_M$, we have measured the
correlation length critical exponent $\nu$ (Fig.~\ref{fig:nu}) and the
skewness of $P(\tE)$ (Fig.~\ref{fig:eskew}).  The results are
completely consistent with those of the Ising model.

Based on the values of the critical exponents and properties of
 $P(\tM,\tE)$, $P(\tM)$ and $P(\tE)$,
we thus conclude that the endpoint
of the SU(2)+Higgs theory is in the universality class of the
3d Ising model.
\smallskip

We have also studied corrections to scaling. While ``symmetric''
corrections to scaling display precisely the exponent inherent for
the 3d Ising model (Sec.~\ref{critpoint}), asymmetric corrections
to scaling, which arise from higher order operators not appearing
in the actions in 
\eqs\nr{isingaction}, \nr{ising}, behave in an unexpected way:
they are quite large at reasonable lattice sizes.
\smallskip

\begin{figure}[t]

\centerline{
\epsfxsize=8cm\epsfbox{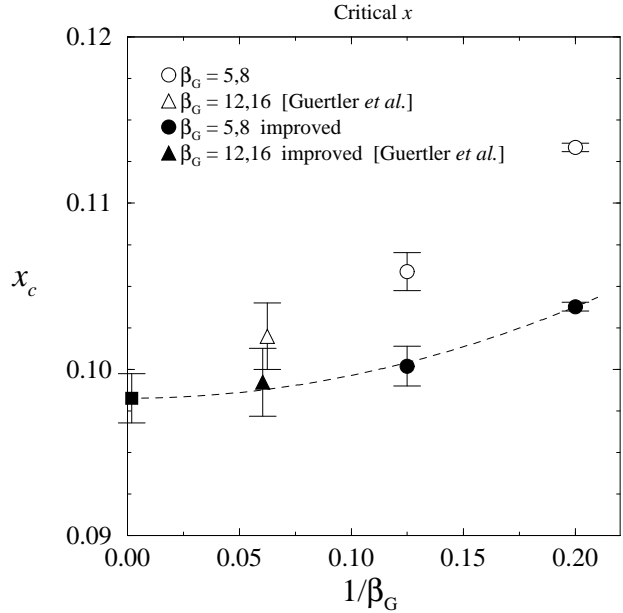}}

\caption[a]{
The infinite volume extrapolations of $x_c$ as a function of $\beta_G$.
G\"urtler et al refers to \cite{gurtler1}.
The location of the endpoint has been determined in \cite{karschnpr},
as well, but the volumes there were somewhat smaller
so that the inclusion
of that datapoint is not meaningful. The improved values   
have been obtained from \eq\nr{eq:moore}.}
\la{xcrit_betag}
\end{figure}

\subsection{The continuum limit of $x_c$}

Finally, let us discuss the continuum limit of $x_c$, the location of
the endpoint, which is a non-universal quantity.  This requires
measurements performed at different lattice spacings; 
for $\beta_G=8$ we used a set of
simulations originally described in \cite{isthere}, and for
$\beta_G=12$ we used the critical coupling measured in
\cite{gurtler1}. 
The results are shown
in Fig.~\ref{xcrit_betag}.
 The improved values for $x_c$ have been obtained
from \eq\nr{eq:moore}.  As the corrections linear in $1/\beta_G$ are
thus removed, a quadratic extrapolation can be made.  The continuum
result is
\be
x_c = 0.0983(15). \la{xcvalue}
\ee
We estimate that the effect of the U(1) group
on $x_c$ is $\lsim 10$\%.
According to the formulas in~\cite{generic}, the value
of $x_c$ in \eq\nr{xcvalue} corresponds
to $m_H=72(2)$ GeV in the SM,
and $x_c=0.11$ would correspond to $m_H=77(2)$ GeV. 
In the MSSM, the same effective
theory as in \eq\nr{action} can be derived, just the relations to
4d parameters are different~\cite{mssm}. Then the value
$x=x_c$ can correspond to many different Higgs masses, depending on
the other parameters of the theory: some examples are shown
in Fig.~\ref{mssm}.

\begin{figure}[tb]

\vspace*{-1cm}

\hspace{1cm}
\epsfxsize=9cm
\centerline{\epsffile{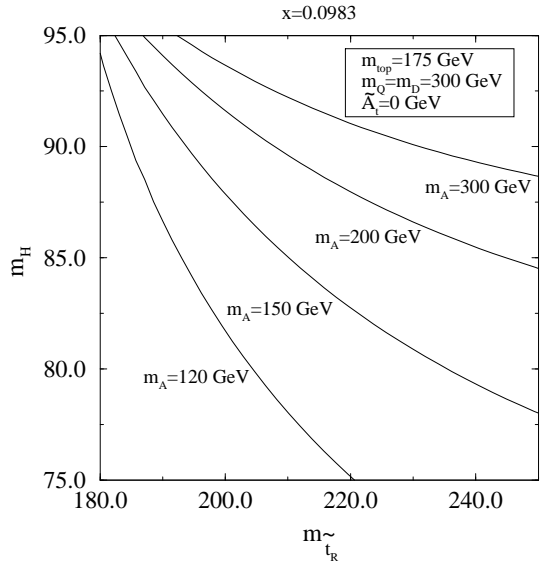}}

\vspace*{-4cm}

\caption[a]{\protect
Examples of parameter values corresponding to
$x=x_c$ in the MSSM~\cite{mssmx}. Here $m_{\tilde t_R}$
is the right-handed stop mass, $m_H$ is
the lightest CP-even Higgs mass and $m_A$ is
the CP-odd Higgs mass. The squark mixing parameters
have been put to zero.}
\la{mssm}
\end{figure}

\section{Conclusions}\la{sec:conclusions}

In this paper, we have shown that the endpoint
of the line of first order transitions in the 3d SU(2)+Higgs theory
is a second order transition in the universality class of the 3d Ising
model. In particular, the measured critical exponents, cumulant
ratios, and probability distributions of these two
theories approach each other with growing lattice size.

To arrive at this conclusion, we have developed a general method to
determine the universality class of a phase transition in a completely
non-perturbative system, utilizing lattice Monte Carlo simulations.
The method can be applied to any system exhibiting critical behaviour,
provided that it is possible to perform Monte Carlo simulations on the
critical point itself.  This includes, e.g., the endpoints of the 1st
order lines in the 3d SU(2)+adjoint Higgs theory (where the line ends
\cite{Hart,su2dr}), the 3d SU(3)+adjoint Higgs theory (where the line turns
into a second order line after a tricritical point) or in the
U(1)+Higgs theory.  On the other hand, the two flavour 4d finite
temperature chiral transition in QCD occurs at the limit $m_q
\rightarrow 0$, which is not directly accessible with standard Monte
Carlo methods.

We have also determined the continuum extrapolation of
an important non-universal quantity, the location
of the endpoint $x_c$.
In the Standard Model with $\sin^2\theta_W=0$, the
resulting value $x_c=0.0983(15)$ corresponds to a physical
Higgs mass $m_H=72(2)$ GeV. While taking $\sin^2\theta_W=0.23$ does not
change the universal properties, the value of $x_c$ may 
grow slightly. However, we do not expect values larger than $x_c=0.11$, 
corresponding to $m_H=77(2)$ GeV. Even this Higgs mass
is already excluded experimentally, and 
thus there is no phase transition in the Standard Model. If low
energy supersymmetry is realized, in contrast, the cosmological
electroweak phase transition can be of the first order for the
Higgs masses allowed at present.
This could lead to important cosmological consequences.

\section*{Acknowledgments}

We gratefully acknowledge useful discussions with
U.M.~Heller and T.~Neuhaus.
The simulations were performed with a Cray T3E at the
Center for Scientific Computing, Finland.
The work of MT was supported by the Russian Foundation
for Basic Research, Grants 96-02-17230, 16670, 16347, 
and that of KK by the
TMR network {\em Finite Temperature Phase Transitions
in Particle Physics}, 
EU contract no.\ FMRX-CT97-0122.

\section*{Appendix}

In this appendix we discuss two points concerning the scalar effective
theory in \eq\nr{isingaction}: 
the relative roles of the linear and cubic terms and
discretization. The cubic term is often used to generate a first order
transition, but here it proves convenient to shift it away.

Consider the theory
\be
S = \int\! d^3x \biggl[
\fr12 \partial_i \phi\partial_i \phi+
h \phi + \fr12 m^2 \phi^2 -\fr13 \delta \phi^3
+ \fr14 \lambda \phi^4 \biggr].
\la{cubicaction}
\ee
Due to superrenormalisability, only the two 2-loop diagrams
\begin{center}
\begin{picture}(240,60)(0,0)
\SetWidth{1.5}
\Line(20,30)(80,30)\CArc(60,30)(20,0,360)
\Line(140,30)(220,30)\CArc(180,30)(20,0,360)
\end{picture}
\end{center}
are logarithmically divergent and lead to the following
renormalisation scale dependence of the mass and
magnetic field terms in the $\msbar$ scheme:
\be
h(\mu)={2\lambda\delta\over16\pi^2}\log{\Lambda_h\over\mu},\quad
m^2(\mu)={-6\lambda^2\over16\pi^2}\log{\Lambda_m\over\mu}.
\la{running}
\ee
The theory now is specified by the four constants
$\lambda,\delta,\Lambda_h,\Lambda_m$ or, equivalently, by
\be
\lambda,\quad x={m^2(\lambda)\over\lambda^2},\quad
y={h(\lambda)\over\lambda^{5/2}},\quad
 z={\delta\over\lambda^{3/2}}.
\ee
However, as there is no $\phi\leftrightarrow-\phi$ symmetry, one
can perform a shift $\phi\to\phi+$constant and choose the
constant
so that the cubic term disappears. This corresponds to the
invariance of the theory under the transformation
$(x,y,z) \to (x+zy/3-2z^3/27,y-z^2/3,0)$. Thus one can from the
outset choose $\delta=0$ in \eq\nr{cubicaction}. The magnetic field
term is then scale invariant (see \eq\nr{running}). However, it is
{\it not} possible to eliminate the linear term: it is anyway
generated by radiative effects according to \eq\nr{running}.

On the lattice the action
corresponding to the theory without
the cubic term becomes (after scaling $a\phi^2\to
\beta_H\phi^2$)
\ba
S&=&-\beta_H\sum_{\bfx,i}\phi(\bfx+i)\phi(\bfx)+
\beta_1\sum_\bfx\phi(\bfx) \nonumber\\
&&+\sum_\bfx\phi^2(\bfx)+{\beta_H^2\over4\beta_G}
\sum_\bfx[\phi^2(\bfx)-1]^2,
\la{scalarlattaction}
\ea
where the lattice couplings $\beta_H,\beta_G,\beta_1$ are
related to $\lambda a,x,y$ by
\ba
\lambda a&=& {1\over\beta_G}, \nn\\
x&=&2\beta_G^2\biggl({1\over\beta_H}-3-{\beta_H\over2\beta_G}
\biggr)+{3\Sigma\over4\pi}\beta_G-{6\over16\pi^2}
\log(6\beta_G+\zeta), \la{phi4relations}\\
y&=&{\beta_1\over\sqrt{\beta_H}}\beta_G^{5/2}.\nn
\ea
Here $\Sigma,\zeta$ are the same as in \eq\nr{y}. 
\eq\nr{scalarlattaction} is a standard scalar lattice action but with
very specific couplings, determined by \eqs\nr{phi4relations}.
If the expectation value of $\phi$
measured with \eq\nr{scalarlattaction} is $\langle\phi_L\rangle$,
then in continuum normalisation,
\be
\langle\phi\rangle/\lambda^{1/2}=\sqrt{\beta_H\beta_G}
\langle\phi_L\rangle.
\ee

If higher order operators are included in \eq\nr{cubicaction},
then the cubic term cannot,
in general, be shifted away any more. It becomes a running parameter
which
is generated radiatively even if shifted away at some scale.


\begin{thebibliography}{99}
\bibitem{mssmx}
This figure is based on the results in the first of~\cite{mssm}.
\end{thebibliography}
\end{document}